\documentclass[hyper,12pt,letterpaper]{JHEP3}
\usepackage{amsmath,amsfonts}
\usepackage{graphicx}
\def\cN{\mathcal{N}}

\def\cN{\mathcal{N}}

\def\inc#1{%
\setbox1\hbox{\includegraphics[scale=.5]{#1}}%
\vcenter{\box1}%
}
\def\inct#1{%
\setbox1\hbox{\includegraphics[scale=.45]{#1}}%
\vcenter{\box1}%
}
\def\inctt#1{%
\setbox1\hbox{\includegraphics[scale=.38]{#1}}%
\vcenter{\box1}%
}

\def\mat#1#2#3#4{\left[\hbox{\footnotesize$\begin{matrix}
#1&#2\\
#3&#4
\end{matrix}$}\right]}
\def\IR{\mathbb{R}}
\def\Z{\relax\ifmmode\mathchoice
{\hbox{\cmss Z\kern-.4em Z}}{\hbox{\cmss Z\kern-.4em Z}} {\lower.9pt\hbox{\cmsss Z\kern-.4em Z}}
{\lower1.2pt\hbox{\cmsss Z\kern-.4em Z}}\else{\cmss Z\kern-.4em Z}\fi}

\def\C{{\bf C}}

\def\cp{{\mathbb{C}}{\bf P}}

\newcommand{\up}{\Upsilon}
\newcommand{\Ga}{\Gamma}

\newcommand{\CF}{{\mathcal F}}

\def\cS{{\mathcal S}}

\def\CF {{\cal F}}

\def\CM {{\cal M}}
\def\CN {{\cal N}}

\def\CT {{\cal T}}

\def\CW {{\cal W}}
\def\CX {{\cal X}}

\def\CZ {{\cal Z}}

\def\p{\partial}

\def\be{\begin{equation}}
\def\ee{\end{equation}}
\def\m{\mu}

\def\bar{\overline}

\def\det{{\rm det}}
\def\tr{{\rm tr}}

\def\Fl{{\CF {\kern -1.2pt \ell} }}

\def\example#1{\bgroup\narrower\footnotefont\baselineskip\footskip\bigbreak
\hrule\medskip\nobreak\noindent {\bf Example}. {\it #1\/}\par\nobreak}
\def\endexample{\medskip\nobreak\hrule\bigbreak\egroup}

\def\btimes{~{{{\lower1pt\hbox{$\square$}} \kern-7.6pt \times}}~}
\def\TT{{\Bbb{T}}}
\def\LL{{\Bbb{L}}}
\def\Weyl{{\cal W}}

\def\C{{\Bbb{C}}}
\def\Z{{\Bbb{Z}}}

\def\m{{\mathfrak m}}

\def\dual#1{{^L\negthinspace #1}}

\title{
Loop and surface~operators
in~$\CN$=\,2~gauge\\
theory and Liouville modular geometry
}

\author{Luis F. Alday,$^1$ Davide Gaiotto,$^1$ Sergei Gukov,$^{2,3}$
Yuji Tachikawa,$^1$
and Herman~Verlinde$^{1,4}$
\\ ~
\\
$^1$ School of Natural Sciences, Institute for Advanced Study,\\
~ Princeton, NJ 08540, USA\\
$^2$ California Institute of Technology, Pasadena, CA 91125, USA \\
$^3$ Department of Physics and Department of Mathematics, \\
~ University of California, Santa Barbara, CA 93106, USA\\
$^4$ Department of Physics, Princeton University,\\
~ Princeton, NJ 08544, USA\\
\\
{\tt alday,dgaiotto,yujitach@ias.edu, gukov@theory.caltech.edu, verlinde@princeton.edu} }

\preprint{CALT-68-2741\\ PUPT-2311}

\abstract{Recently, a duality between Liouville theory and four dimensional ${\cal N}=2$ gauge theory
has been uncovered by some of the authors. We consider the role of
extended objects in gauge theory, surface operators and line operators, under this correspondence.  We map such objects to specific
operators in Liouville theory. We employ this connection to compute the expectation value of general
supersymmetric 't Hooft-Wilson line operators in a variety of ${\cal N}=2$ gauge theories.}

\begin{document}

\addtolength{\parskip}{.7mm}
\addtolength{\baselineskip}{.2mm}
\addtolength{\abovedisplayskip}{.8mm}
\addtolength{\belowdisplayskip}{.8mm}

\bibliographystyle{utphys}

\section{Introduction}

Recently, a rich class of four-dimensional (4d) $\CN\,$= 2
superconformal gauge theories was identified as the infrared
fixed point of the 4d theory obtained by compactifying the
six-dimensional (2,0) superconformal theory of $A_{N-1}$ type
on a general Riemann surface $C$ with punctures
\cite{Gaiotto:2009hg}\cite{Gaiotto:2009we}. In the IR limit, the 4d gauge theory
data do not depend on the scale factor of the 2d metric on
$C$. The space of coupling constants of this class of 4d superconformal gauge theories
can be identified with Teichm\"uller space $\CT_{g,n}$,
the universal covering space of the moduli space $\CM_{g,n}$ of
complex structures of punctured Riemann surfaces. Moreover, via
the six-dimensional perspective, S-duality naturally arises as
the geometric invariance under the action of  the mapping class
group $\Gamma_{g,n}$, the group of large diffeomorphisms acting
on $C$ that leave its complex structure fixed. The space of
physically inequivalent superconformal gauge theories thus takes the form of the
quotient \be \CM_{g,n} = \raisebox{2pt}{$\CT_{g,n}$}{\bigr/}
\raisebox{-2pt}{$\Gamma_{g,n}$}\, . \ee

A practical subset among this class of $\CN=2$ gauge theories, that is most
accessible to quantitative computations, is obtained by
compactifying the six-dimensional (2,0) theory of type $A_1$. In
this case, the superconformal gauge theory admits a weakly coupled Lagrangian
description, whenever the compactification surface $C$
degenerates into a set of three-punctured spheres (also known
as `trinions'  or `pairs of pants') glued together via thin
tubes. The weakly coupled theory takes the form of a generalized
quiver gauge theory, where each tube corresponds to an $SU(2)$
gauge group factor, and each trinion represents a matter
multiplet, transforming as a trifundamental under the three
adjacent $SU(2)$ factors, and each puncture to an ungauged
$SU(2)$ flavor group. %\footnote{
The simplest examples are
$\CN\,$= 4 and $\CN\,$= $2^*$ super Yang-Mills (SYM) theory with gauge group
$SU(2)$, corresponding to the torus with zero and one puncture,
 and the $\CN=2$ $SU(2)$ gauge theory with $N_f=4$ flavors, corresponding to the
 four-punctured sphere. %}
The geometric operations that  connect different ways of assembling the
same Riemann surface become identified
with S-duality transformations, which relate different Lagrangian descriptions of the same
theory. The dictionary between the dual descriptions involves a generalization of
electric-magnetic duality, that exchanges the role of electric and magnetic
observables such as the Wilson and 't Hooft loop operators.

The generalized quiver diagrams, which specify the perturbative
limits of the $\CN=2$ gauge theory  associated with a Riemann surface
$C$, look identical to the trivalent graphs that are used to
label the conformal blocks of a two-dimensional (2d) CFT~on ~$C$. It
is then natural to suspect that there may exist a direct
correspondence between S-duality operations of the
4d superconformal gauge theory and  modular transformations of conformal blocks in some suitable 2d CFT.

This intuition was recently made precise in \cite{Alday:2009aq},
where it was shown that the Nekrasov instanton partition function
of the generalized quiver gauge theory on $\IR^4$ is identical to the conformal block
(specified by the corresponding trivalent graph)
in Liouville conformal field theory.
In this correspondence, the Liouville momenta at the marked points specify
the masses of the flavor multiplets, while the momenta in the intermediate
channels are identified as the Coulomb branch parameters.
The central charge of the Liouville CFT
is determined by the value of  two
deformation parameters $\epsilon_1$ and $\epsilon_2$,
which can be identified with the coordinates
on the Lie algebra of the rotation group
$SO(2)_1 \times SO(2)_2 \subset SO(4)$ acting on the $\IR^4$.
Furthermore, it was found that the full Liouville
correlation function, which takes the form of the integral
of the absolute value squared of conformal blocks, naturally arises as
the partition function of the 4d $\CN=2$ gauge theory defined on $S^4$ \cite{Pestun:2007rz}.

These remarkable relations allow for a multi-pronged analysis
of the properties of this class of theories. A  useful general
strategy is as follows.
\vspace{-1mm}

\begin{itemize}
\item Pick a class of observables $\{ {\cal O} \}_{6d}$
in the six-dimensional $A_1$ theory on $C$.
\vspace{-1mm}
\item On the Coulomb branch, the $A_1$ theory reduces
to the free abelian theory for a single M5-brane wrapped
on the Seiberg-Witten curve $\Sigma$, a double cover of~$C$.
The flow of $\{ {\cal O} \}_{6d}$  towards the
IR can be easily followed, giving rise to a class of observables $\{ {\cal O} \}_{\mathfrak{u}(1)}$
of the 4d abelian Seiberg-Witten gauge theory.
\vspace{-1mm}
\item The result becomes more
useful if one can identify the meaning of the observables
directly in the 4d  generalized quiver gauge
theories. A natural way to accomplish that is to employ the
perspective of brane constructions in type IIA string theory.
This defines a new incarnation of the observables, $\{ {\cal O} \}_{4d}$.
\vspace{-1mm}
\item The relation to the 6d observables $\{ {\cal O} \}_{6d}$
provides $\{ {\cal O} \}_{4d}$ with a manifest behavior
under S-duality, and a map from $\{ {\cal O} \}_{4d}$
to the IR observables $\{ {\cal O} \}_{\mathfrak{u}(1)}$.

\vspace{-1mm}
\item Finally, one can seek a Liouville theory manifestation of these
observables, $\{ {\cal O} \}_{2d}$. The powerful methods developed in
the context of 2d conformal field theory can be applied to the
computation of expectation values of $\{ {\cal O} \}_{4d}$
on the four sphere, or in the $\epsilon$-deformed background on~$\IR^4$.

\end{itemize}

\vspace{-1mm}

In this paper we will employ this general strategy to study
three natural classes of observables in the 4d gauge theory
(i) general Wilson-'t Hooft line operators, (ii) surface
operators and (iii) line operators bound to surface operators.
In particular we will illustrate how to compute the expectation
value of these operators by using Liouville CFT technology.

\bigskip

\subsection{Surface, line and point operators}

The six-dimensional perspective gives useful guidance in
identifying and relating the various  gauge theory observables.
The  (2,0) theory of type $A_1$ arises as the infrared limit of
the world-volume theory of a stack of two coincident M-theory
five-branes (together with a free 6d theory describing the
center-of-mass motion). Each M5-brane contains a two-form
potential $B$ with self-dual three-form field strength.
An M2-brane can attach to an M5-brane via an open boundary,
that sweeps out a 2d surface $\cS$.
It is a source for $B$.
The different ways of embedding $\cS$ inside the 6d space-time
$C \times {\mathbb R}^4$ give rise to three different classes of gauge theory observables:
(1) surface operators,
(2) line or loop operators, and (3) point or `vertex' operators:

\begin{figure}[t] \centering \includegraphics[width=5.4in]{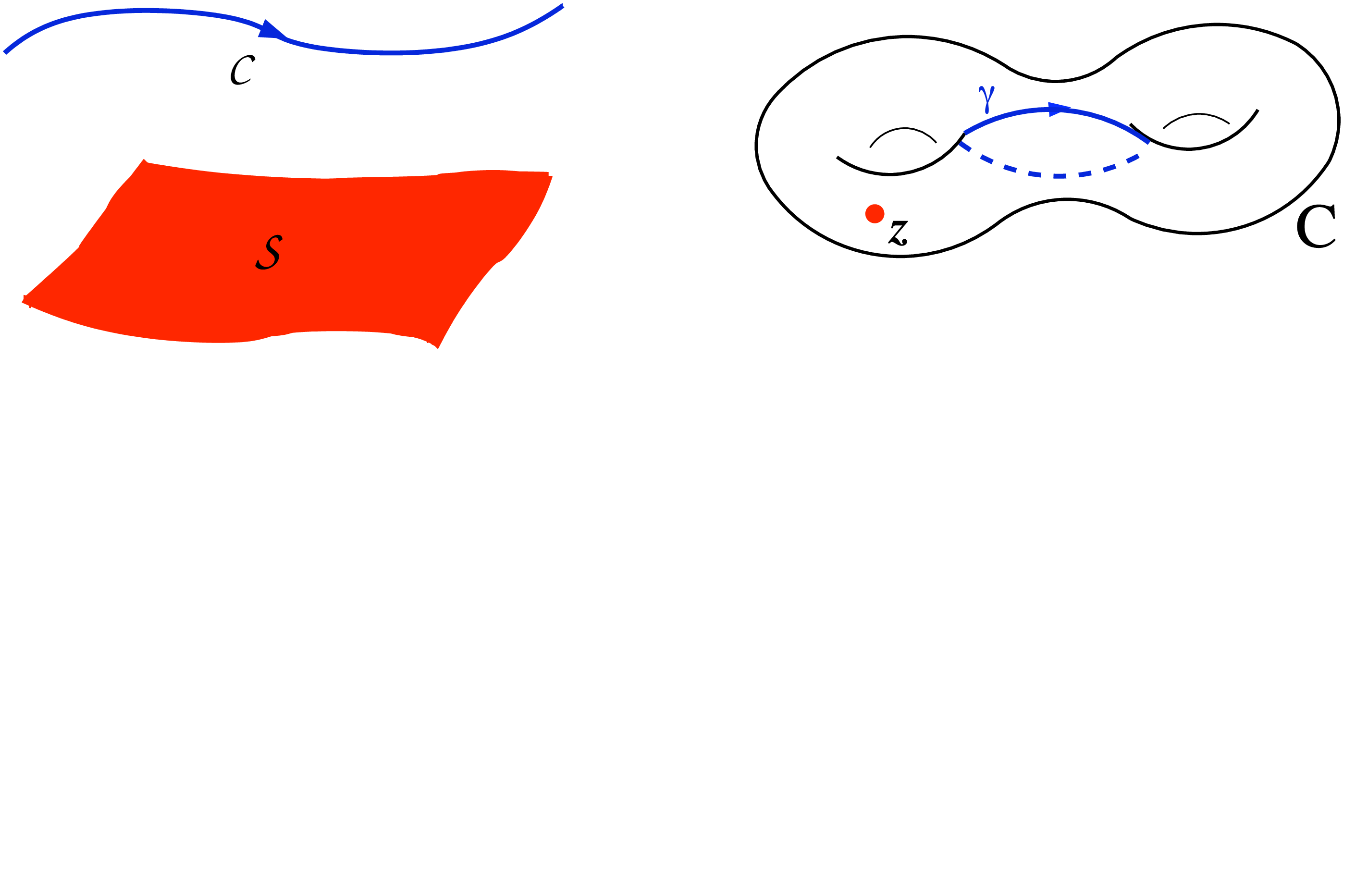} \caption{Surface operators
are supported on a surface $\cS$ in ${\mathbb R}^4$ (shown on the left part of the figure)
and are localized at a point $z$~in~$C$ (on the right).
Similarly, line operators extend along an (open or closed) curve ${\mathcal C}$ in ${\mathbb R}^4$
and wrap a 1-cycle $\gamma$~in~$C$.} \label{observables}\end{figure}

\begin{enumerate}

\item  The surface operators are defined by considering an
    M2 boundary surface $\cS$ to be embedded\footnote{Although
    in this paper we mainly take $\cS=\IR^2$,
    in the topological version of the theory one might consider more
    general space-time 4-manifolds $M$ and embedded surfaces $\cS \subset M$,
    {\it cf.} \cite{Witten:1988xj,Kronheimer:1993,Gukov:2006jk,Gukov:2007zz,Witten:2007td}.}
    in the 4d space-time ${\mathbb R}^4$ and localized as a point $z$
    on $C$. In $\CN\,$= 4 SYM theory, the surface
    operators are identified \cite{Gukov:2006jk} as
    operators that create a singular vortex by allowing for
    a suitable singular boundary condition on the gauge
    and scalar fields along $\cS$.
%A surface operator breaks the gauge group $G$ down to
%a subgroup $\LL \subset G$, the so-called Levi subgroup \cite{Gukov:2006jk}.
For the most elementary class of surface operators, %i.e. those with next-to-maximal Levi subgroup $\LL$,
the vortex singularity
    is parametrized by two real parameters $\alpha$
    and~$\eta$; here $\alpha$ is the magnetic flux through
    the singular vortex and $\eta$ is a suitable 2d
    theta-angle. Both are naturally defined as periodic
    variables; from the M-theory point of view, they
    parametrize the location $z$ of the surface operator on $C = T^2$.

As we explain below, a similar class of half-BPS surface operators
can be defined in $\CN\,$ = 2 quiver gauge theories of interest.
Moreover, for the most elementary class of such operators,
the parameters ($\alpha, \eta$) associated
to the different $SU(2)$ gauge group factors can be glued together to specify a
single location $z$ on the punctured Riemann surface~$C$.

\item The line or loop operators are represented by
    M2-brane boundaries that wrap a one-cycle $\gamma$ on $C$,
    and extend along an infinite line or closed loop
    ${\cal C}$ in ${\mathbb R}^4$. In the perturbative regime,
    where the surface $C$ decomposes into thin tubes sewed
    together via trinions, the loops labeled by the
    one-cycles  around the thin tubes represent fundamental
    Wilson lines of the corresponding $SU(2)$ gauge groups.
    General Wilson-'t Hooft line operators can be thought
    of as the coupling of the gauge theory to the worldline
    of a dyonic point charge. The spectrum of possible
    Wilson-'t Hooft loops in the generalized quiver gauge
    theory is labeled by the set of closed
    non-selfintersecting paths on $C$, up to homotopy
    \cite{Drukker:2009tz}. As explained in \cite{Drukker:2009tz},
    in a given weakly coupled description in terms of gauge theory
    with gauge group $G = SU(2)^{3g-3+n}$, this set has the physically expected form.

Line operators can act on surface operators, when the
worldline ${\cal C}$ of the former is embedded inside the
worldsheet $\cS$ of the latter. The line operator then
creates a discontinuity along ${\cal C}$ in the parameters
$(\alpha,\eta)$ of the surface operator, generated by
transporting its location $z$ on $C$ by the corresponding
closed path on $C$. Intuitively, we can think of this
discontinuity as the effect of the generalized Dirac string
of the dyonic point particle.

\begin{figure}[t] \centering \includegraphics[width=4.5in]{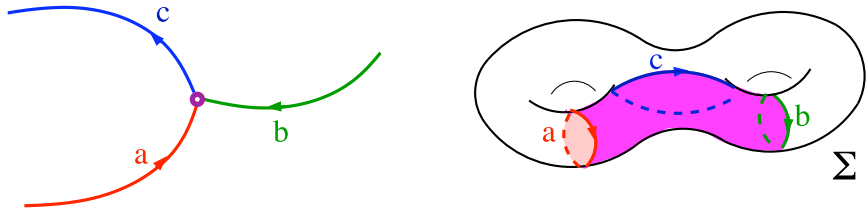} \caption{A point or `vertex' operator may form a junction between several line operators. On $C$, it spans
an open region bounded by the (non-intersecting) one cycles associated with the line operators that meet at the junction. } \label{vertex}\end{figure}

\item Point or `vertex' operators may form a junction
    between several line operators. On $C$, they span on
    open region bounded by the (non-intersecting) one
    cycles associated with the line operators that meet at
    the junction. In the simplest case, when the boundary consists of three Wilson line operators
    in three adjacent gauge group factors, the point operator represents a point charge transforming in the 
    corresponding trifundamental representation.

\end{enumerate}

\noindent
In this paper we will focus our attention on the surface and loop operators,
and leave the study of the point operators for future work.

\bigskip

\subsection{Computation strategy}

We now summarize the basic strategy of our calculation of the expectation value of general
Wilson-'t Hooft line operators on $\IR^4$ and $S^4$.  Although the validity of the actual computation
does not rely on any unverified assumptions,  it turns out that we can gain some useful
geometric intuition by first stating the following conjecture:

\bigskip

\noindent ${}$~~\parbox{14.5cm} {\it The expectation value  in
the  $\CN\, $= 2 gauge theory of an elementary surface operator,
specified by its position $z$ on $C$,
is equal to the Liouville CFT correlation function with the added insertion
of a degenerate primary operator $\Phi_{2,1}(z)=e^{- (b/2) \phi(z)}$.}

\bigskip

\noindent
Although the complete proof of this conjecture goes beyond the scope of the present paper,
in Sections 2 and 3 we present several pieces of evidence that support this proposed identification.
For now, however, we will adopt it as a working hypothesis,
that will help us formulate a practical procedure for computing
the expectation values of Wilson-'t Hooft loops by means of the Liouville CFT correlation functions.

\begin{figure}[t] \centering \includegraphics[width=3.3in]{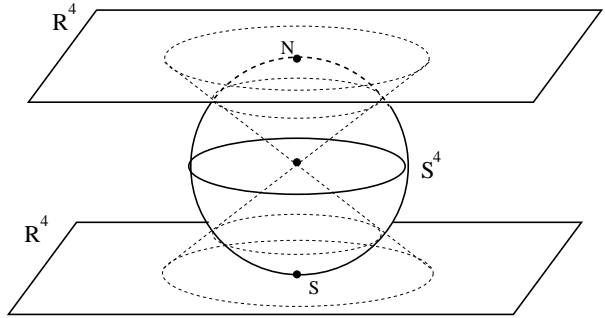} \caption{The hemispherical stereographic projection of
$S^4$ onto two copies of $\IR^4$. It reflects the factorization of the instanton sum on $S^4$ into two ``chiral'' halves,
given by the $\IR^4$ contribution of instantons localized near the north and south pole. Surface operators on $S^4$ similarly factorize into a two ``open'' surface operators, a north and a south half, glued together at the equator.} \label{stereo}\end{figure}

Let us state the conjecture a bit more precisely.
As shown in \cite{Alday:2009aq}, the Nekrasov partition function
on ${\mathbb R}^4$ is equal to a Liouville conformal block, i.e. a {\it  chiral half} of the full Liouville
correlation function, while the partition function on $S^4$ takes the form of an integral of the absolute value
squared of a conformal block. So it is natural to identify the division of $S^4$ into the northern and southern
hemispheres with the chiral decomposition of the Liouville CFT correlation functions into ``left-moving"
and ``right-moving" chiral halves. To make this somewhat more concrete,
imagine choosing hemispherical stereographic coordinates on $S^4$ as indicated in fig \ref{stereo}.
The upper and lower halves of $S^4$ are projected on two copies of $\IR^4$.
We parametrize each $\IR^4 \cong \C^2$ by two complex coordinates $(w_1,w_2)$ and $(\tilde{w}_1,\tilde{w}_2)$,
such that the north and south pole of the $S^4$ project to the origin of the corresponding $\IR^4 \cong \C^2$.

Now imagine adding a single elementary surface operator,
inserted, say, on the lower copy of $\IR^4$. In the gauge theory
set-up of \cite{Nekrasov:2002qd} and \cite{Pestun:2007rz},
there are two natural locations for the surface operators,
namely $w_1=0$  {and}  $w_2 = 0.$ Both locations are invariant under
the $U(1)$ %_1 \times U(1)_2$
rotation symmetry used in the localization
of the gauge theory path integral, which acts as
\begin{align}\label{torusact}
(w_1,w_2) \mapsto (e^{2\pi i \epsilon_1} w_1,%~ w_2) \cr
%U(1)_2 &:  (w_1,w_2) \mapsto (w_1,
~ e^{2\pi i \epsilon_2} w_2)
\end{align}
As we shall argue below, the expectation value of the simplest type
of such surface operators located at $w_1=0$ corresponds to the insertion,
inside the Liouville CFT conformal block,
of a degenerate chiral vertex operator $\Phi_{2,1}(z)$,
while the same type of surface operators located at $w_2=0$
corresponds to the insertion of the chiral operator $\Phi_{1,2}(z)$,
which is the quantum version of the Liouville exponential $e^{-\phi(x)/(2b)}$.
Indeed, these two types of surface operators are related
by the symmetry~$\epsilon_1~\leftrightarrow~\epsilon_2$.
According to the dictionary of \cite{Alday:2009aq},
in the Liouville theory it corresponds to switching the roles of $b$ and $1/b$,
which indeed relates the degenerate chiral vertex operators $\Phi_{2,1}(z)$ and $\Phi_{1,2}(z)$.
Note that the conformal block is {\it multi-valued} as a function of
the position $z\in C$. This multi-valuedness arises because
this class of surface operators on $\IR^4$ has an open
boundary at infinity~\cite{Gukov:2006jk}.

Via the hemispherical stereographic projection, the surface
operator on $S^4$ can be thought of as the result of gluing
together two ``open''  surface operators, one acting on the
south copy of $\IR^4$ and one acting on the north copy of
$\IR^4$. We conjecture that the expectation value of the
surface operator on $S^4$ is given by inserting a non-chiral
vertex operator inside the non-chiral Liouville correlation
function. Note that the non-chiral correlation function of
$\Phi_{2,1}$ is {\it single valued} as a function of  $z \in C$,
which is as one would expect for surface operators that do not
have any open boundary. The factorization of the non-chiral
operator into left- and right-moving chiral vertex operators
amounts to splitting the closed surface operator into two ``open''  halves.

\begin{figure}[t] \centering \includegraphics[width=5.7in]{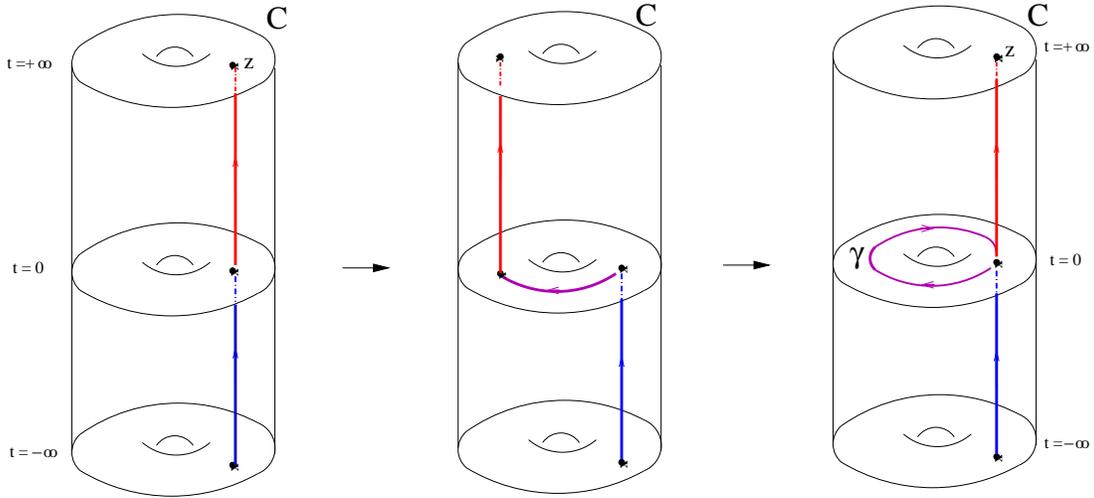} \caption{A Wilson-'t Hooft loop  is labeled by a closed path $\gamma$ on $C$, and a surface operator is specified by a location $z$ on $C$. When the loop acts
on a surface operator on $S^4$, it shifts the relative location of the upper- and lower-half
via a monodromy operation associated with the closed path $\gamma$.
The vertical direction indicates a `time coordinate' $t$ on $S^4$, defined such that the equator gets mapped to $t=0$
and the north and south pole to $t=\pm \infty$.} \label{monodromy}\end{figure}

Next, consider a Wilson-'t Hooft loop labeled by a closed path
$\gamma$ on $C$,  acting on a surface operator on $S^4$. For
concreteness, we take the surface operator to be located at
$w_1\,$= 0, and the loop operator to act within the equator of
the $S^4$. The loop operator splits the surface operator into
two open halves, glued together via a prescribed discontinuity
in the parameters $\alpha$ and  $\eta$ of the singular vortex,
i.e. via a jump in the location $z \in C$.   Since the two
sides correspond to the two chiral halves of the degenerate field
$\Phi_{2,1}$, the discontinuity  amounts to a relative shift in
the location $z$ of the left and right chiral vertex operators
by a full monodromy around $\gamma$. We can thus visualize the
action of the Wilson-'t Hooft loop as performing a monodromy
operation, in which one of the chiral vertex operators is
transported along the closed path $\gamma$. This procedure is
illustrated in fig \ref{monodromy}.

Finally, to get a Wilson-'t Hooft line in isolation, one can
start by viewing it as the result of annihilating two identical
surface operators, i.e. both are at the same location on $S^4$
and on $C$, except that one of the two
has a discontinuity as a result of acting with the given
 loop operator at the equator. Via the geometric
visualization of the discontinuity as drawn in fig
\ref{monodromy}, we arrive at the identification of  the
Wilson-'t Hooft loop with the following familiar CFT monodromy
operation:\footnote{ In the gauge theory, the four steps
correspond to: \
 (i) insert a ``trivial''
surface operator at $w_1\,$=\;0,\;
(ii)  split it  into a pair of conjugate
surface operators, each specified by the same parameter $z$ on $C$,\;
(iii) act with a  loop operator on one of the two surface operators,\;
(iv) let the two conjugate
surface operators annihilate each other, leaving behind a bulk loop operator.}

\vspace{-1mm}

\begin{enumerate}
\item Insert the identity operator $\mathbf{1}$ inside the Liouville correlation function.
\vspace{-1mm}
\item Write  $\mathbf{1}$  as the result of fusing two degenerate Liouville operators $\Phi_{2,1}(z)$,
via their operator product expansion.
\vspace{-1mm}
\item Transport the chiral half of one of the two operators along the closed non-self-intersecting path $\gamma$
that labels the Wilson-'t Hooft operator.
\vspace{-1mm}
\item Reconstitute the local operator $\Phi_{2,1}$ (by recombining the two chiral halves),
and re-fuse  the two degenerate fields together into identity via the OPE.
\end{enumerate}

\vspace{-1mm}

The above monodromy procedure
 was  introduced in the context of
rational CFT by E.~Verlinde \cite{Verlinde:1988sn},  and played a key role in the
derivation of the relation between modular transformations and
the fusion algebra.  It defines a linear operator, that acts
non-trivially on the space of conformal blocks. To explicitly
perform the various steps, one needs to know the modular
properties of the conformal blocks under basic moves, known as
fusion and braiding. Liouville field  theory is a non-rational
CFT, but  its conformal blocks have rather similar modular
properties as in rational CFT, except that the labels are  continuous
rather than discrete \cite{Teschner:2003en}, \cite{Teschner:2001rv}.
In particular, the fusion and braiding
matrices are known explicitly, and satisfy the necessary
polynomial consistency relations. This knowledge is sufficient
for us to turn the above four step procedure into a
straightforward computation of the expectation value of the
Wilson-'t Hooft line operators.

\bigskip

\subsection{Organization}

The rest of this paper is organized as follows. In Section 2, we set up a
semi-classical dictionary between the $\CN\,$=2 gauge theory
and Liouville theory, based on the asymptotics of the Nekrasov
partition function and the identification between the
expectation value of the Liouville  energy-momentum tensor and
the quadratic differential describing the Seiberg-Witten curve. We pay special
attention to the semiclassical behavior of the monodromies
of the degenerate Liouville field $\Phi_{2,1}$.

In Section 3, we first recall the definition of the surface operators in the gauge
theory, and provide an M-theory realization of them.
We then present a semi-classical argument that supports
their identification with the insertion of the degenerate fields in Liouville theory.

In Section 4, we consider the action of Wilson-'t Hooft loop operators on
surface operators, and relate their expectation values to the monodromy of $\Phi_{2,1}$ in the full
quantum Liouville theory. As an application, we discuss how S-duality between the Wilson and
't Hooft loops follows from elementary properties of CFT conformal blocks.
In Section 5 we consider the Wilson-'t Hooft loop operators in the bulk, and use the
recipe outlined above to compute their expectation value for the specific examples of
$\CN=2^*$ and $N_f=4$ SYM theory. We briefly discuss their relation
to observables on quantized Teichm\"uller space. 
We end with some
concluding comments on open problems and future directions in
Section 6.

In Appendix A and B, we have collected some useful
facts about Liouville modular geometry, the form of the
relevant fusion and braiding matrices and the relations among
them. In Appendix C, we present an explicit calculation of the
semi-classical limit of a degenerate operator insertion.
Finally in Appendix D, we discuss the issue of self-intersecting paths
on the Seiberg-Witten curve.

\section{Semi-classical Liouville/gauge theory correspondence}

In this section, we give a short overview of the semi-classical limit of Liouville CFT and
its correspondence with the Seiberg-Witten solution of the class of $\CN\,$=\, 2 gauge theories
introduced in  \cite{Gaiotto:2009we}. We then use this correspondence to study the semi-classical
monodromies  of the Liouville degenerate field $\Phi_{2,1}$.

\subsection{Seiberg-Witten curve from Liouville}

The IR dynamics of undeformed $\CN\,$=\;2 gauge theories on $\IR^4$ is completely characterized by
the classical Seiberg-Witten (SW) curve. For our class of theories, the SW curve is given by the double cover $\Sigma$
of the Riemann surface $C$, specified in terms of a quadratic differential $\phi_2(z)$ defined on $C$, as
\begin{equation}
x^2 = \phi_2(z)\, .
\end{equation}
$\phi_2(z)$ has double poles at the $n$
marked points, whose coefficients encode the mass parameters
$m_i$ of the gauge theory. The space of quadratic differentials
with double poles of fixed coefficients is an affine space of
dimension $3g-3+n$. This is also the dimension of the Coulomb
branch.  The Coulomb branch moduli $a_i$ of the field theory
are identified with periods of the SW differential
$\lambda_{SW} = x dz$ around a complete set of non-intersecting
one-cycles $A_i$ on $\Sigma$
\begin{equation}
\label{aperiods}
\frac{1}{2\pi i}\oint_{A_i} \! x\,dz = a_i \, .
\end{equation}
The periods of the SW differential around the dual cycles $B_i$
on $\Sigma$ specify a dual set of parameters
\begin{equation}
\label{bperiods}
\frac{1}{2\pi i}\oint_{B^i}\! x\, dz  = a_D^i \, .
\end{equation}
The magnetic  parameters $a_D^i$ are not independent from the $a_i$, but determined via
\begin{equation}
\label{adual}
a_D^i = \frac{1}{4\pi i} \frac{\partial {\cal F}}{\partial a_i} \, ,
\end{equation}
where ${\cal F}$ is the SW prepotential, which is an analytic function of the $3g-3+n$ coupling constants $\tau_i$ and
Coulomb branch parameters $a_i$. % associated with the $SU(2)$ gauge group factors.

In the perturbative limit, there is a canonical choice of $A^i$
cycles which project to a complete set of mutually
non-intersecting closed paths in the Riemann surface $C$, that
surround the thin tubes that characterize the $SU(2)$ gauge
group factors of the generalized quiver gauge theory. The
reader is warned this choice ceases to be canonical as soon as
one moves away from the perturbative limit. The homology
lattice of the SW curve is subject to all sort of interesting
monodromies as one varies $\phi_2$. At a generic point in the
Coulomb branch, there is no preferred choice of a set of
special coordinates $(a_i, a^i_D)$.

To help compute the instanton partition sums of $\CN\,$=\,2 gauge theory, Nekrasov considered a
deformation of the Lagrangian by two parameters $\epsilon_1$ and $\epsilon_2$, both with the dimension of mass,
that specify a certain $SO(4)$ rotation and some non-commutative modification of the space-time $\IR^4$.
This $\epsilon$ deformation breaks the
translational symmetry and effectively places the functional integral on a compact space-time: the full
partition function on $\IR^4$ is just a finite number $\CZ_{4d}$, which depends meromorphically on the coupling
constants and Coulomb branch parameters. Interestingly, $\CZ_{4d}$ coincides with a
 conformal block of the Liouville CFT defined on the base curve $C$, with conformal fields $V_{m_k}(z_k)$
 placed at the $n$ punctures \cite{Alday:2009aq}:
\begin{eqnarray}
\label{conformalblok}
\CZ_{4d} =\Bigl \langle  V_{m_1}(z_1) \cdots V_{m_n}(z_n)\Bigr\rangle_{\{ a_i\} }
\end{eqnarray}
Here, in our notation for the conformal block, we leave
implicit the choice of pants decomposition of the Riemann
surface $C$. Both sides of this equality are given as a
perturbative expansion in the instanton factors $q_i = e^{2 \pi
i \tau_i}$ of the $SU(2)$ gauge groups, which are identified
with the parameters of the ``plumbing fixture'' used to join
the various pairs of pants. For example, when the base curve $C$ is a sphere,
$q_i$ give the cross ratios of the coordinates $z_i$ of the insertions.\footnote{
There is a certain degree of arbitrariness in the precise
definition of conformal blocks. Pairs of pants are glued together by a local coordinate transformation
$z_1 z_2 = q$.  The exact parameterization of the complex structure moduli space by the $q_i$
depends  on the precise choice of a local coordinate at each puncture. Fortunately, the
integration kernels implementing S-duality do not depend on the $q_i$, and are thus insensitive to this choice.
The instanton partition function suffers of similar arbitrariness, in the sense of some regularization scheme dependence. The ambiguity did not manifest itself in the explicit examples of \cite{Alday:2009aq},
possibly because of an underlying brane construction.}

\newcommand{\spc}{\hspace{.5pt}}

In a sense, that we will make more precise in what follows, the Nekrasov deformation amounts to a ``quantization''
of the space of Coulomb branch parameters, that specify the SW differential of our class of theories.
In accordance with this interpretation, we write\footnote{Note that this does {\em not} become the standard practice $\epsilon_1=\hbar$, $\epsilon_2=-\hbar$ at $b=1$. }
\be
\epsilon_1 = b\spc \hbar, \qquad \qquad
\epsilon_2 = \frac{\hbar}{b}\, .
\ee
Here $\hbar$ defines some mass scale, relative to which we will measure all other mass parameters.
The parameter $b$ is related to the central charge $c$ of the Liouville CFT via $c=1+6Q^2$ with $Q=b+\frac{1}{b}$.
The fields inserted at the $n$ marked points have Liouville momentum $\frac{m_k}{\hbar}$ and
conformal dimensions $\Delta_k = \frac{m_k}{\hbar} (Q-\frac{m_k}{\hbar})$, where $m_k$ is the mass parameter for the $SU(2)$ flavor group associated to the $k$-th puncture. The  primary field 
propagating in intermediate channels 
is given by $e^{\alpha_i \varphi(z)}$ with
\be
\label{momenta}
\alpha_i = \frac{Q}{2} + \frac{a_i}{\hbar}
\ee where $a_i$ is the Coulomb branch parameter,
In other words, $a_i$ specifies the Liouville momentum in the channel.\footnote{In the following, we will refer to the exponent $a_i$ as the Liouville momentum.}
This operator has conformal dimension $\Delta_i =
(\frac{a_i}{\hbar} + \frac{Q}{2} )(\frac{Q}{2}-\frac{a_i}{\hbar})$.
The SW curve and associated prepotential $\CF(a)$ emerges from the Nekrasov partition function in the ``semiclassical limit'' $\epsilon_{1,2} \ll  a_i,m_i$, or in 2d terminology, the $\hbar \to 0$ limit where all Liouville momenta become large:
\be
\label{semic}
\log {\cal Z}_{4d} \; \simeq \; -\frac{\mbox{\small 1}}{\mbox{\small $\hbar^2$}} \, \CF(a) \; + \; \ldots
\ee

Here the canonical choice of $A$-cycles is playing a hidden role.
As both sides of \eqref{semic} are defined by power series
in the $q_i$, the logarithm and the $\hbar \to 0$ limit should
be taken term by term in the $q_i$ expansion.  The important
monodromies of $(a_i, a^i_D)$ in the Coulomb branch are
completely invisible to the $q_i$ expansion: each term is a
rational function of the $a_i$.

As was observed in \cite{Alday:2009aq},
the quadratic differential $\phi_2(z)$ that specifies the SW curve can be recovered in the
semiclassical limit $\hbar \to 0$ from the Liouville CFT, by considering the expectation value of the 2d energy momentum tensor
\be
\label{tofz}
\Bigl\langle T(z) V_{m_1}(z_1) \cdots V_{m_n}(z_n)\Bigr\rangle_{\{a_i\}} \to\;  -\frac{1}{\hbar^2}\, \phi_2(z)\,
\Bigl\langle  V_{m_1}(z_1) \cdots V_{m_n}(z_n)\Bigr\rangle_{\{ a_i\}}
\ee
The quadratic differential $\phi_2(z)$ defined this way has
double poles at $z_k$ with coefficient given by $\hbar^2$ times
the conformal dimension $h_k$, which in the semi-classical
regime coincides with the squared mass parameter $m_k^2$.
Similarly, it is not hard to verify that the definition
(\ref{aperiods}) of the $a_i$ parameters with the electric
periods of the SW differential $x dz = \sqrt{\phi_2(z)}  dz$
around the  $A_i$ cycles, perfectly matches with the
identification (\ref{momenta}) with the intermediate Liouville
momenta $\alpha_i$. Again, the match is to be understood
term-by-term in the $q_i$ expansion.

It was shown by Pestun \cite{Pestun:2007rz} that the instanton partition function
of the undeformed gauge theory on $S^4$ is given by the
integral over the Coulomb branch parameters $a_i$ of the
absolute value squared of the $\IR^4$ partition function, with
equal deformation parameters $\epsilon_1= \epsilon_2 = 1/R$,
where $R$ is the radius of $S^4$:
\be \CZ_{S^4} = \int \! d a_i
\, \left|\Bigl \langle  V_{m_1}(z_1) \cdots
V_{m_n}(z_n)\Bigr\rangle_{ \{ a_i\} }\right|^2\, . \ee
This expression coincides with the partition function of the full
non-chiral Liouville field theory. Since non-chiral CFT
partition functions are invariant under modular
transformations, this observation makes explicit that the $S^4$
partition function $\CZ_{S^4}$ is S-duality {\it invariant}.
In contrast, Nekrasov's partition function on $\IR^4$
transform non-trivially under S-duality.

Indeed the conformal blocks labeled by different trivalent graphs
can be considered as different delta-function normalizable bases
of the same Hilbert space, labeled by the continuous parameters $\alpha_i$.
The change of basis involves integration against an intricate kernel,
which does not depend on the $q_i$.
If we denote the choice of trivalent graph for the quiver, or the conformal block,
as ${\cal G}$, we can write schematically
\begin{eqnarray}
\label{S-duality}
\CZ_{4d}^{\cal G}(a_i, m_k;\tau_i) =\int \! da'_i\, \CZ_{4d}^{\cal G'}(a'_i, m_k;\tau'_i) \, {\cal K}(a_i, a'_i,m_k) ,
\end{eqnarray}
where the $\tau_i$ are the complex structure moduli of the surface in the new basis.

\subsection{Monodromies of the degenerate field $\Phi_{2,1}$}

To gain more insight, it is useful to introduce the insertion of a degenerate local Liouville operator $\Phi_{2,1}$.
As mentioned in the introduction, and explained in more detail in Section 3,
we propose that this operator insertion corresponds to the gauge theory
partition sum in the presence of an elementary surface operator. % (of the next-to-maximal Levi type).

Let us consider the properties of $\Phi_{2,1}$ in the
semi-classical limit. The degenerate field $\Phi_{2,1}$ can be
viewed as the operator with Liouville momentum equal to $-b/2$.
It satisfies the relation $(L_{-1}^2 + \spc b^2 L_{-2})
\Phi_{2,1} = 0$, which implies that, when inserted in any
correlation function, it satisfies a differential equation of
the form
\begin{equation}\label{degendiff}
 \partial_z^2 \Phi_{2,1}(z)= - b^2 :\!T(z) \Phi_{2,1}(z)\!:
\end{equation}
Here the normal ordering amounts to subtracting the double and single pole singularity as $T(z)$ approaches $\Phi_{2,1}(z)$.

For a general surface $C$ with $n$ punctures, the above
differential equation has a large space of solutions, which one
would like to identify with the space of conformal blocks with
a degenerate insertion.\footnote{ The identification is true, but with an
important caveat. The null vector $(L_{-1}^2 + \spc b^2 L_{-2})
\Phi_{2,1}$ decouples  from correlation functions, but
surprisingly does not decouple automatically from conformal
blocks as well, unless one imposes ``by hand'' the degenerate
fusion rule: the Liouville momenta on the two sides of the
degenerate insertion must differ by $\pm \frac{b}{2}$. We will
assume this constraint whenever we talk about conformal blocks
with one or more degenerate insertions. A few more details are
given in appendix \ref{app:degfusion}. } The choice of sign in $\pm
\frac{b}{2}$ corresponds to the two solutions of the second
order differential equation (\ref{degendiff}).

Since the conformal dimension of $\Phi_{2,1}$ is fixed, and thus remains finite as $\hbar \to 0$, in the semi-classical
regime one is allowed to replace $T(z)$ by its expectation value (\ref{tofz}). The semiclassical analysis
of (\ref{degendiff}) thus is reduced to the WKB analysis of a holomorphic Schr\"odinger equation.

\newcommand{\PPsi}{{\cal Z}} %{\mbox{\large $\Psi$}}

Consider the conformal block with a  degenerate field insertion.
\begin{equation}
\PPsi( a_i \, ;\, z) \, = \;  \Bigl\langle \Phi_{2,1}(z)\, V_{m_1}(z_1) \cdots V_{m_n}(z_n) \Bigr\rangle_{\{a_i\}}
\ee
The insertion modifies the semi-classical limit
 (\ref{semic}) at subleading order, to
\be
\label{wkb}
\PPsi(a_i ; z)\, \sim \, \exp \Bigl(\, -\frac{\CF(a_i)}{\hbar^2} +
\frac{b \CW(a_i,z)}{\hbar}+\ldots \, \Bigr)\, .
\end{equation} A basic WKB argument, combining (\ref{degendiff}) and (\ref{tofz}),  shows that
\begin{equation}
\label{difw}
(\partial_z \CW)^2 = \phi_2(z) = x(z)^2,
\end{equation}
hence $\CW$ is (plus or minus) the integral
of the SW differential along some path to the point $z$, starting at some reference point $z_*$:
\begin{equation}
\label{wpm}
\CW_\pm(z) = \pm  \int_{z_*}^z \!\! x \, dz
\end{equation}
The choice of sign in (\ref{wpm}) corresponds to the two-fold degeneracy in the space of conformal blocks
with a degenerate insertion.\footnote{As we will see in Section 3, in the gauge theory, the two fold degeneracy arises because the
IR surface operators associated with a given gauge group factor have two degenerate vacua.} We will denote the two
WKB solutions by $\PPsi_\pm(a_i;z)$.

{}
Since the SW differential has non-vanishing periods (\ref{aperiods}) and (\ref{bperiods}),  (\ref{wkb}) and (\ref{wpm}) tell us
that $\PPsi_\pm(a_i;z)$ is a multi-valued function of the position $z$ of the degenerate field. As we will see later (and as detailed in appendix \ref{app:degfusion}),
in the full quantum CFT this multi-valuedness is implemented via so-called fusion and braiding matrices, which
in this case are given by $2\times 2$ matrices that relate the doublets of conformal blocks
with the degenerate field inserted at different locations.
The transport of $z$ along paths in the Riemann surface
is implemented by the composition of a certain number of these matrices.
We will denote the resulting transport operator along a path $\gamma$ as $\CM_\gamma$.

As a crude first step towards finding the semiclassical behavior of
 $\CM_\gamma$, we could simply look at the monodromy of each WKB wavefunction.
This monodromy depends on the periods of the SW differential
along the lift of $\gamma$ to the SW curve.\footnote{In the following intuitive argument,
we will temporarily ignore some important structure associated to the
fact that the same homotopy class in the base  curve $C$ lifts to a multitude of possible
homology classes in the SW curve. Still, the naive reasoning is rather instructive.}
The monodromy of the WKB wavefunctions around the $A$-cycle $A_i$ on $\Sigma$ is given by a simple
phase factor, determined by the corresponding Coulomb branch
parameter\footnote{Here $z+A_i$ is a schematic notation for
moving the position $z$ on $C$ along the cycle $A_i$.}
\be
\label{amonodromy} \PPsi_\pm(a_i ; z+ A_j) = \exp\bigl(\pm
\mbox{\large $ \frac{2\pi i  b}{\hbar}$}\, a_j\bigr)\,
\PPsi_\pm\bigl(a_i ; z\bigr). \ee
This behavior is as expected from standard CFT arguments: transporting a degenerate field around
a certain leg of the conformal block produces a simple phase factor
$e^{2 \pi i (\Delta_\alpha + \Delta_{2,1} - \Delta_{\alpha \pm {b}/{2}})}$.
This agrees at the leading order with (\ref{amonodromy}).

The $B$-cycle monodromy, on the other hand, takes the form
\begin{eqnarray}
\label{bmonodromy}
\PPsi_\pm(a_i \, ; z+ B^j) &=& \exp\bigl(\pm \mbox{\large $\frac{2\pi i b}{\hbar}$}\, a_D^j\bigr)\, \PPsi_\pm\bigl(a_i ; z\bigr)\, .%\nonumber
\end{eqnarray}
Via eqn. (\ref{adual}) and working to leading order in $\hbar$,
we see that the prefactor in (\ref{bmonodromy}) can be naturally absorbed
via a quantized shift in the Coulomb branch parameter associated with the dual $A$-cycle:
\be
\label{cmonodromy}
\PPsi_\pm(a_i \, ; z+ B^j) \, = \, \PPsi_\pm\bigl(a_i \pm \mbox{\large $\frac{b \hbar}{2}$} \, \delta_i^j\, ;\, z\bigr)\, .%\nonumber
\ee
Hence we see that the $B$ cycle monodromies may lead to shifts
in the $a_i$ parameters by multiples of $\hbar b/2$. We will confirm this fact via a more precise quantum treatment in Section
\ref{sec:line}.
\smallskip

As explained in the Introduction, the above monodromy operations represent the
action of Wilson loops (for the $A_i$ monodromy) and 't Hooft loops (for the $B^j$ monodromy) on a surface operator in the gauge theory.  The above naive semi-classical expressions for these monodromies, while incomplete,
already give some useful first hints at what general structure we should expect for the full answer.

%\begin{enumerate}

First, we see that the conformal blocks with fixed $a_i$ parameters naturally
form an eigenbasis of the $A_i$ monodromies of $\Phi_{2,1}$.
The $B_i$ monodromies, on the other hand, act non-trivially on the eigenlabels $a_i$.
In the gauge theory, this corresponds to the fact that the instanton partition function (in the presence of a surface operator)
on $\IR^4$ is an eigenfunction of the Wilson loop operator, while the 't Hooft loop operator acts on the Coulomb branch parameters via quantized shifts.\footnote{A priori, it may look somewhat surprising that the 't Hooft loops can change the
Coulomb branch parameters, and do not commute with the Wilson line operators.
However, as noted earlier, the $\epsilon$ deformation effectively makes the space compact. Thus
a localized operator may be capable of changing the vevs $a_i$. Secondly,  loop operators that act on
a surface operator can be ordered in `time'; hence it is meaningful to talk about commutators between loop operators.}
 S-duality can thus be thought of as a change
of eigen basis from a set of `electric' loop operators to some dual set of  `magnetic'  loop operators.

Note further that the expectation value of a surface operator on $S^4$, which is expressed as the integral over the Coulomb branch parameters of the absolute
valued squared of the instanton sum, is a single-valued function
of~$z$: the $A$ monodromies are phase factors that do not affect the norm squared, while the shifts in $a$
generated by the $B$ monodromies can be absorbed in a redefinition of the integration variables. This distinction between
$\IR^4$ and $S^4$ expectation values is related to the fact that surface operators on $\IR^4$ are open (and thus
 may produce boundary terms upon partial integration), while on $S^4$ they are closed.

%{({\it iii}\/)  Eqns (\ref{bmonodromy}) and (\ref{cmonodromy}) suggest that the electric Coulomb branch parameters
%$a_i$ and the magnetic dual parameters $a_D^j$ can be thought of as conjugate quantum mechanical
%variables, with a commutation relation of the form $ [\, a_i, a_D^j\, ] \simeq \frac{i \hbar^2}{4\pi}
% \delta_i{}^j$.  This naive observation can be turned into a more precise statement,
%originally conjectured in \cite{Verlinde:1989ua} and recently proven by Teschner \cite{Teschner:2005bz},
%that Liouville conformal blocks on a Riemann surface $C$ can be thought of as states in the Hilbert space
% obtained by quantizing the Teichm\"uller space $\CT_{g,n}$ of $C$. Via this perspective,  the S-duality group arises
 %as the group of unitary transformations that implement canonical transformation on the phase space $\CT_{g,n}$.  }

%\end{enumerate}

\smallskip

The above comments are all meant as intuitive
expectations, based on a somewhat crude semi-classical arguments. The
WKB approximation can be conducted in a  rather more precise
way, following the approach of  \cite{Gaiotto:2009hg}. A crucial step in \cite{Gaiotto:2009hg} was a careful WKB analysis of a certain
differential equation involving the same quadratic differential $\phi_2(z)$ as we have here.
This method can be applied with minor
modifications to the holomorphic Schr\"odinger equation based on $\phi_2(z)$.
The trick is to re-express the transport matrices
$\CM_\gamma$ for the differential problem as a linear combination
of certain quantities $\CX_{\tilde \gamma}$ , for which the
naive WKB approximation along the path $\tilde \gamma$ in the
SW curve is correct. As detailed in appendix A of \cite{Gaiotto:2009hg},
$\CM_\gamma$ is a linear combination, with integer coefficients,
of $\CX_{\tilde \gamma}$, where the index  $\tilde \gamma$ runs
over various possible lifts of $\gamma$ to the SW curve. At
different values of the parameters, different terms in the sum
may be dominant in the semi-classical limit. Moreover, the
integer coefficients which determine which $\tilde \gamma$ is
actually present in the sum are subject to discontinuous jumps
as a function of the parameters.
Only in a fixed perturbative limit, the naive WKB approximation around the $A^i$ cycles is valid.
This is a rather degenerate case of the analysis in  \cite{Gaiotto:2009hg}, where a maximal set of ``closed WKB curves'' emerges.

The analogy between our setup and the setup of \cite{Gaiotto:2009hg} is
clearly not coincidental. The relation between  $M_\gamma$ and
$\CX_{\tilde \gamma}$ in \cite{Gaiotto:2009hg} represents the IR behavior of
the same general class of line operators in the gauge theory as
we consider here. 

In Sections \ref{sec:line} we will compute the full quantum expression for the
monodromies (\ref{amonodromy}) and (\ref{cmonodromy}) in Liouville CFT, which will provide the
exact gauge theory expectation values of the Wilson and 't Hooft line operators. The results will
confirm the basic intuitive picture presented above.

\section{Surface operators  in $\CN=2$ Gauge Theories}

In this section we discuss %in more detail
a simple brane realization of
half-BPS surface operators in $\CN=2$ gauge theory 
and argue that their counterparts
in the low energy effective theory are labeled by points on the SW curve.
Here we consider general $SU(N)$ gauge groups
because our construction works equally well for any $N$.  
We will restrict our attention to $SU(2)$ 
in Sec.~\ref{sec:line-on-surface} and \ref{sec:line}, 
and compare gauge theory data with Liouville theory data.

As we will see, the brane construction shows that the twisted superpotential
of the 2d theory on the surface operator is given by
an integral along an open path on the SW curve $\Sigma$,
reproducing the formulae  (\ref{wkb}) and (\ref{wpm}),
thus supporting our identification between surface operators in the gauge theory side
and insertions of degenerate operators in the Liouville side.
We will also comment on how we can derive these results from the
instanton counting in the presence of a surface operator.

Before we start, we should mention
a relationship between our consideration of the surface operators
and the analysis of quantum vortices in  the Higgs phase of  $\cN=2$ theories
presented in the review \cite{Tong:2008qd} and references therein.
There, supersymmetric Nielsen-Olsen type vortices were considered
in the maximal Higgs branch of a specific $\cN=2$ theory, and the quantum dynamics
of the zero modes living on the vortices was studied. They found
a relation similar in spirit as \eqref{difw}, although they could only probe a very special point
on the Coulomb branch, namely the root of the maximal Higgs branch.
There is an obvious, sharp distinction between these vortices and our surface operators: 
the former are dynamical excitations of the theory, whereas the latter are operator insertions. 
Nevertheless, the two results are not completely unrelated. It is possible to
consider a setup where an $\cN=2$ theory sits at the bottom of an IR flow initiated in a larger theory by 
a judicious Higgs branch expectation value. Vortex strings in the larger theory will flow in the far IR to 
surface operators: the magnetic fluxes of the low-energy $U(1)$ gauge fields in the core of the vortex 
are squeezed to delta functions, and the tension of the strings goes to infinity.
In a similar spirit, one can establish a relation between our surface operators 
in $\cN=2$ field theories, and the D-strings employed by \cite{Ooguri:1999bv} in $\cN=2$ string theory compactifications to give a physical interpretation to 
the refined open topological string amplitudes\cite{Gukov:2004hz,Iqbal:2007ii}. 

\medskip

\subsection{Half-BPS surface operators}
\smallskip

First let us recall the ultraviolet definition of surface operators.
We are interested in half-BPS surface operators in $\CN=2$
gauge theories. The super-Poincar\'e subgroup preserving the
surface operator corresponds to $\CN = (2,2)$ supersymmetry in two
dimensions. More specifically, if we denote the two sets of
4d supercharges as $Q^\pm_\alpha$, where $\pm$ denotes
the eigenvalue of the Cartan generator of $SU(2)_R$,
the surface operator preserves a left moving half of $Q^+_\alpha$ and
a right moving part of $Q^-_\alpha$. This is motivated by the
fact that we consider (mass deformation of) 4d superconformal
theories, and the natural subgroup of the superconformal group
which preserves a surface operator is
\be
\label{supergp}
SU(1,1|1)_L \times SU(1,1|1)_R \times U(1)_e \, \subset \, SU(2,2|2).
\ee

Four-dimensional multiplets restricted to the surface operator
can be packaged into 2d superfields,
useful to describe the couplings to the 2d defect.
Different 2d supermultiplets can be identified
with the help of the extra $U(1)_e$ factor in (\ref{supergp}),
which commutes with the 2d superconformal group.
The $U(1)_e$ is a linear combination of the Cartan generator of
$SU(2)_R$ and of the rotations in the plane transverse to the
surface operator.

Abelian vector multiplets in four dimensions restricted to the
surface operator yield a twisted chiral multiplet of charge
$0$ under $U(1)_e$.
Every such multiplet contains the 2d part of the field strength,
together with the vector multiplet scalars.
Twisted superpotential terms integrated over the surface operator
will play a role which is quite parallel to the role of the
prepotential in the 4d theory, as they are functions of the
Coulomb branch vevs. Expanding in component, they give rise to
couplings to the abelian 4d magnetic and electric fluxes across
the surface operator.

For a surface operator that breaks the gauge group $G$ down to
a subgroup $\LL \subset G$ (the so-called Levi subgroup \cite{Gukov:2006jk})
one can introduce a 2d Fayet-Iliopoulos (FI) term of the form $t \int_{\cS} C$
for each abelian factor in $\LL$.
A simple example corresponds to the next-to-maximal $\LL$,
{\it e.g.} $\LL = U(N-1) \times U(1)$ (or $SU(N-1) \times U(1)$)
in a theory with gauge group $G=U(N)$ (resp. $SU(N)$).
In this case, there is only one FI parameter $t$,
which can be conveniently written as $t = \eta + \tau \alpha$
in terms of real parameters $\alpha$ and $\eta$
that have a simple interpretation in gauge theory \cite{Gukov:2006jk}.
Namely, the ``magnetic'' parameter $\alpha$ defines a singularity for the gauge field:
\begin{equation}
\label{asing}
A = \alpha d \theta + \cdots,
\end{equation}
where $x^2 + i x^3 = r e^{i \theta}$ is a local complex coordinate,
normal to the surface $\cS \subset M$,
and the dots stand for less singular terms.
Note, in order to obey the supersymmetry equations,
the parameter $\alpha$ must take values in the $\LL$-invariant part of
$\frak t$, the Lie algebra of the maximal torus $\TT$ of $G$.

On the other hand,
the ``electric'' parameter $\eta$ enters the path integral through the phase factor
\begin{equation}
\label{etaphase}
\exp \left( i \eta \cdot \m \right)
\end{equation}
where
\begin{equation}
%{}\qquad\qquad\qquad\qquad\qquad\qquad
\m
= {1 \over 2\pi} \int_{\cS} F %\qquad\qquad\qquad\qquad {\rm (``monopole~number")}
\end{equation}
measures the magnetic charge of the gauge bundle $E$ restricted to $\cS$.
The monopole number $\m$ takes values in the $\Weyl_{\LL}$-invariant part of the cocharacter lattice,
$\Lambda_{{\rm cochar}}$, which we denote as $\Lambda_{\LL}$.
The lattice $\Lambda_{\LL}$ is isomorphic to the second cohomology group
of the flag manifold $G/\LL$, a fact that will be useful to us later.
Therefore,
\begin{equation}\label{mlattice}\m ~\in~ \Lambda_{\LL} \cong H_2 (G/\LL ; \Z)\end{equation}
and the character $\eta$ of the abelian magnetic charges $\m$
takes values in ${\rm Hom} (\Lambda_{\LL},U(1))$, which is
precisely the $\Weyl_{\LL}$-invariant part of $\dual{\TT}$.

The ``classical'' twisted superpotential coupling on the
surface operator which is associated to these surface operators
is simply $(\eta + \tau \alpha) a$, where $a$ is the
superpartner of $F$, the restriction of the Coulomb scalar
field to $\cS$. On the Coulomb branch of the non-abelian gauge
theory, the twisted superpotential will evolve into an
effective twisted superpotential $\CW$, a non-trivial function
of the abelian Coulomb branch parameters. We propose that in
general the effective twisted superpotential, much like the
effective prepotential, is computable in terms of the
SW curve $\Sigma$. In particular, we claim that it
is given by the integral of the SW differential
along an open path (starting at some reference point $p_*$) on
the SW curve,
\begin{equation} \label{eq:super}
\CW = \int_{p_*}^p \lambda
\end{equation}
The endpoint $p$ of the path provides an IR parameterization of
surface operators.

Notice that in the IR abelian gauge theory, the superpotential
is a function of the Coulomb branch parameters $a^i$ of the
abelian gauge fields. The couplings to electric and magnetic
fluxes $t_i = \eta_i + \tau_{ij} \alpha^j$ live naturally in
the Jacobian variety of the SW curve.%
\footnote{More properly the Prym variety. For example in the
$A_1$ case, the derivatives of the SW differential $\lambda$
with respects of the parameters $u_i$ in $\phi_2=\lambda^2$
produce holomorphic differentials $\omega_i =
%\frac{1}{2\lambda} \frac{\partial \phi_2}{\partial u_i}$
\frac{\partial \lambda}{\partial u_i}$
which are odd under the involution $\lambda \to -\lambda$ of the
ramified cover $\Sigma\to C$. }
Because the partial derivatives of the SW differential are,
by definition, the holomorphic differentials $\omega_i$ on the SW curve,
the map
\begin{equation}
t_i = \frac{\partial \CW}{\partial a^i} = \int_{p_*}^p \frac{\partial \lambda}{\partial a^i} = \int_{p_*}^p \omega_i
\end{equation}
coincides with the Abel-Jacobi map from a Riemann surface to its Jacobian.

\subsection{Surface operators from M2-branes}

Let us now study how these surface operators arise
in terms of a surface operator in the six
dimensional $(2,0)$ $A_{N-1}$ theory on a Riemann surface $C$.
In terms of M5-branes, this is a setup where $N$ M5-branes wrap
$C \times \IR^{4} \times \{ {\rm pt} \}$ in $T^*C \times \IR^{4}\times \IR^3$. The
$SU(2)_R$ R-symmetry rotates the transverse $\IR^3$,
while the $U(1)$ R-symmetry acts on the fiber of $T^*C$.
The surface operator represents the endpoint of an M2-brane,
stretched to infinity along a specific direction in $\IR^3$.
%This selects a Cartan subalgebra of $SU(2)_R$.
Therefore,
the resulting surface operator is naturally labeled by a point $z$ in $C$,
when all the M5-branes are coincident and thus the theory is at the origin
of the Coulomb branch.

In the Coulomb branch of the theory, the M5-branes merge into
a single M5-brane wrapping the SW curve $\Sigma$,
an $N$-ramified cover of $C$ in $T^*C$ defined by an equation
\begin{equation}
x^N = \sum_{i=2}^N \phi_i(z) x^{n-i}
\end{equation}
Here $\phi_i(z) dz^i$ are degree $i$ differentials on $C$. The
normalizable deformations of the $\phi_i(z) dz^i$ correspond to
the Coulomb branch parameters. The SW differential
is $\lambda = x dz$. The M2-brane ends on the SW
curve at a point $p=(x,z)$.

As the abelian theory on a single M5-brane is well understood,
we can understand directly the coupling of the surface operator
to the fluxes of the 4d abelian gauge theory. The 4d fluxes are
the components of the self-dual three form field strength on
the M5-brane along the harmonic one forms on the SW curve. The
surface operator couples to the self-dual three form field
strength in a standard way: pick a three chain bounded by the
surface operator $p \times \IR^{2}$ and a reference surface
$p_* \times \IR^{2}$ and integrate the three form field
strength along it. We reproduce the desired
\begin{equation}
t_i = \int_{p_*}^p \omega_i
\end{equation}
and from that the twisted superpotential.

How should we understand the relation between the UV label $z$
and the IR label $p=(x,z)$? The degrees of freedom living at
the UV surface operator appear to have $N$ distinct vacua in
the IR. We would like to interpret the SW equation as a chiral
ring relation for a twisted chiral superfield $x$, capturing
the dynamics of such degrees of freedom. At least locally, if
we consider a small variation of the 2d coupling $z \to z + \delta z$,
it is natural to consider $\delta z$ as an FI
parameter for the twisted chiral superfield $x$.

The twisted chiral superfield $x$ resembles closely the
generator of the quantum chiral ring of a $\cp^{N-1}$ sigma model.
This is not a coincidence.
The basic surface operators in the $SU(N)$ gauge theory,
which break the gauge symmetry to the subgroup
$\LL \cong SU(N-1) \times U(1)$,
have a natural relation to a $\cp^{N-1}$ sigma model:
one can always re-instate full gauge symmetry on the surface operator
by introducing a compensator field living in
$SU(N)/(SU(N-1) \times U(1)) = \cp^{N-1}$.
This compensator field may well become dynamical in the IR.

A weakly coupled $SU(N)$ gauge group in four dimensions arises from
the M5-brane theory whenever a tube in the Riemann surface
$C$ is close to degeneration, {\it i.e.} becomes long and thin.
In this limit the $(2,0)$ theory along the tube can be reduced to
a 5d Yang-Mills theory in a segment, and then to a weakly coupled 4d
$SU(N)$ gauge theory. The gauge coupling $\tau$ is the modular
parameter of the tube. If the M2-brane is attached to (one of)
the M5-branes in the long tube region, it will clearly produce
a defect in the 4d $SU(N)$ gauge theory which breaks $SU(N)$ to
$SU(N-1) \times U(1)$.

We can be more precise. As we reduce from the $(2,0)$ theory on
a long, thin tube to a weakly coupled $SU(N)$ 5d Yang-Mills
theory on a long segment, the M2-brane surface operator
descends to a D2-brane surface operator, represented in the 5d
theory by a 't Hooft monopole operator of minimal charge. The
position of the original puncture on the M-theory circle is
encoded in the angle $\eta$ coupled to the magnetic flux
integrated over the surface operator. By supersymmetry, the
holomorphic coordinate $t$ along the tube must coincide with
the holomorphic combination $\eta + \tau \alpha$. We still have to fix a
reference point, $t=0$, that will be discussed below.

To summarize, we conclude that the definition of standard
surface operators in $\CN=4$ SYM theory can be easily extended
to surface operators in $\CN=2$ theories in the weak coupling
regime, provided that the punctures are well inside the tubes of
the Riemann surface. As the puncture moves through pair of pants
from one tube to another, the corresponding surface operator
must undergo some interesting 2d duality transformation.

To understand better the detailed structure of the surface operators,
we will follow a standard route \cite{Gaiotto:2009we}:
we will first focus on a subclass of theories,
the conformal linear quivers of unitary groups,
which have a brane realization in IIA theory \cite{Witten:1997sc}.

\begin{figure}[t] \centering \includegraphics[width=5.0in]{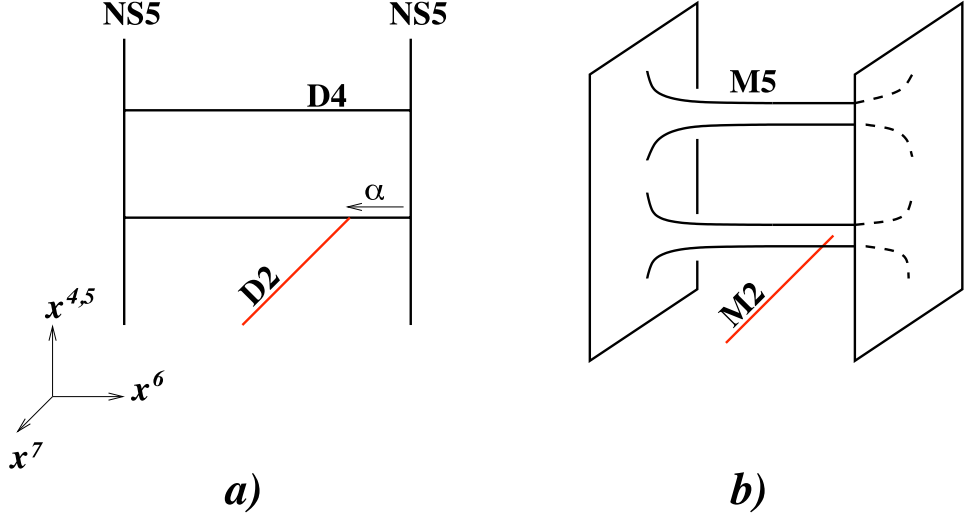} \caption{The brane construction of $\CN=2$ super Yang-Mills theory
with a half-BPS surface operator in type IIA string theory $(a)$ and its M-theory lift $(b)$.\label{branefig}} \end{figure}

\subsection{Brane construction in type IIA}

Let us consider a stack of $N$ D4-branes intersecting $n$ NS5-branes.
We take the NS5-branes to be along the directions $x^0$, $x^1$, $\ldots$, $x^5$,
and D4-branes to be along the directions $x^0$, $x^1$, $\ldots$, $x^3$, and $x^6$.
This setup realizes a conformal linear quiver of $n-1$ $SU(N)$ groups,
with $N$ fundamental hypers at each end.
In M-theory, it lifts to a brane configuration
which we identify with the $A_{N-1}$ theory ``compactified'' on a cylinder, with $n$ simple defects.

To produce transverse, semi-infinite M2-branes in the M-theory
setup we need transverse, semi-infinite D2-branes in the IIA setup.
They should preserve half of the remaining supersymmetry
of the D4 and NS5 brane system. We choose the D2-brane
worldvolume to be along the directions $x^0$, $x^1$, and $x^7$,
as in Figure~\ref{branefig} $a$. Below we summarize the worldvolume
directions of various branes in the resulting configuration:
\begin{align}\label{iiabranes}
\hbox{NS5} &: \quad 012345 \cr
\hbox{D4}  &: \quad 0123~~~6 \cr
%D6  &: \quad 0123~~~~789 \cr
\hbox{NS5$'$} &: \quad 01~~~45~~~~~89 \cr
\hbox{D2}  &: \quad 01~~~~~~~~7
\end{align}
where we included a new kind of the five-brane, denoted as NS5$'$,
with worldvolume along the directions $x^0$, $x^1$, $x^4$, $x^5$, $x^8$, and $x^9$.
The NS5$'$-brane preserves the same part of the 4d $\CN=2$
supersymmetry as the D2-brane and is useful for identifying
the half-BPS surface operator represented by the D2-brane.

\begin{figure}[t] \centering
\includegraphics[width=5.0in]{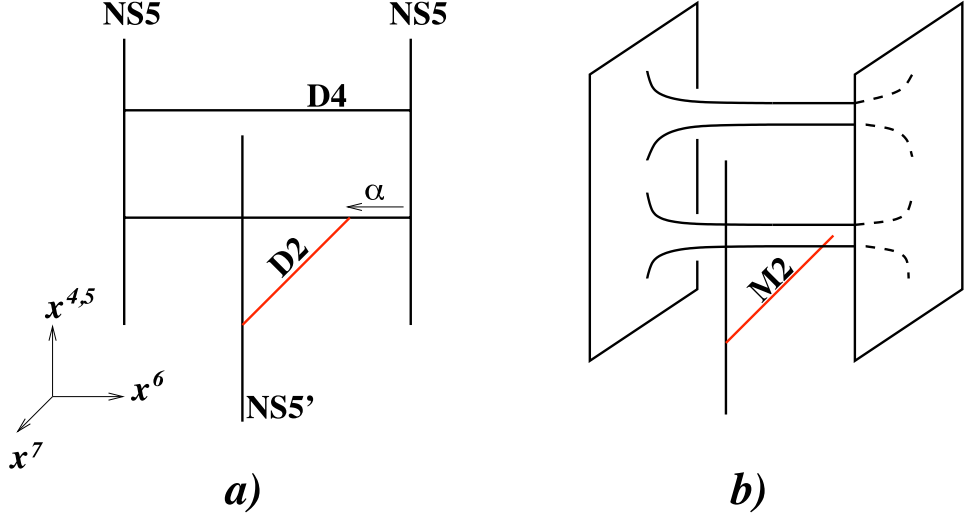}
\caption{\label{dbranefig}
$(a)$ The brane construction of $\CN=2$ super Yang-Mills theory
with a half-BPS surface operator (shown on Figure~\protect\ref{branefig} $a$)
where we introduced an extra NS5$'$-brane.
Now the D2-brane can end on the NS5$'$-brane, thus, having a finite extent in the $x^7$ direction.
$(b)$ The M-theory lift of the type IIA brane configuration on part $(a)$ of the figure.} \end{figure}

%\subsubsection{Effective Theory on the D2-brane}

In the presence of the NS5$'$-brane, the D2-brane can have
a finite extent in the $x^7$ direction
by stretching between the NS5$'$-brane on the one end,
and the original system of D4 and NS5 branes, on the other,
as illustrated on Figure~\ref{dbranefig} $a$.
When the D2-brane has finite extent in the $x^7$ direction,
its worldvolume theory is effectively a 2d
$U(1)$ gauge theory with the coupling constant
\begin{equation}\label{dtwocoupling}
{1 \over e^2} = {\ell_s \Delta x^7 \over g_s} \end{equation}
In particular, the original brane configuration on Figure~\ref{branefig} $a$
can be recovered in the limit $\Delta x^7 \to \infty$,
which corresponds to the weak coupling limit of the D2-brane theory.

To be precise, it is convenient to start by attaching
the D2-brane to one of the NS5-branes. The Neumann boundary
conditions on the D2-brane worldvolume theory allow for a
simple dimensional reduction to a 2d gauge theory.
The effective theory on the D2-brane is $\CN=(2,2)$
supersymmetric gauge theory with gauge group $U(1)$
and a certain matter content, which is easy to read of from the brane
construction on Figure~\ref{dbranefig} $a$. Specifically, we have the
following $\CN=(2,2)$ theory in two dimensions
(with space-time coordinates $x^0$ and $x^1$):
\begin{equation}\label{dtwotheory}
{\bf D2~theory:} \quad U(1) {\rm ~with~} N {\rm ~chiral~multiplets~of~charge~}1{\rm ~and~}N{\rm ~of~charge~}-1 \nonumber\end{equation}
Indeed, up to a simple change of coordinates,
this setup is related to the brane system considered in \cite{Hanany:1997vm}
that engineers $\CN=(2,2)$ 2d abelian gauge theory
with chiral multiplets of charge $+1,-1$.
In the D2-brane theory, the boundary conditions corresponding
to the NS5 and NS5$'$ branes project out all massless string modes,
except for a $\CN=(2,2)$ vector multiplet.
Indeed, since the NS5-brane is localized in the directions $x^6$, $\ldots$, $x^9$,
and since the NS5$'$-brane is localized in the directions $x^2$, $x^3$, $x^6$, $x^7$,
the D2-brane can only move along the directions $x^4$ and $x^5$
(which are common to both the NS5 and NS5$'$ brane).
These two modes combine into a complex scalar field $\sigma$
on the D2-brane worldvolume,
$$
{x^4 + i x^5 \over \ell_s^2} \Bigg|_{{\rm D2}} = \sigma
$$
which can be identified with a complex scalar
in the 2d $\CN=(2,2)$ vector multiplet
(equivalently, twisted chiral multiplet).

Once we have D2, NS5, and NS5$'$ branes,
incorporating the D4-branes does not break supersymmetry further.
The D2-D4 open string states give rise to charged chiral multiplets
(one for every D4-brane) resulting in the effective theory (\ref{dtwotheory}).
Note that, in the D2-brane theory, the vevs $a_i$ of the 4d
adjoint scalar field play the role of twisted mass parameters
and the $SU(N)$ gauge symmetry of the 4d gauge theory
on the D4-branes plays the role of the flavor symmetry.
For generic values of $a_i$ the $SU(N)$ flavor symmetry is broken to a subgroup $U(1)^{N-1}$.
We get a set of $N$ chiral multiplets of charge $+1$ from the D4-branes ending
on the left of the NS5-brane,
and a set of  $N$ chiral multiplets of charge $-1$ from the D4-branes ending
on the right of the NS5-brane.

These 2d fields couple in a standard way to
the 4d gauge fields arising from the D4-branes,
and also couple (via cubic superpotential) to the bifundamental
adjoint hypermultiplets coming from the D4-D4 strings
(stretched between the two sets of D4-branes).
Giving expectation values to the bifundamental fields corresponds
to reconnecting the D4-branes and separating them from the NS5-brane;
this operation is known to give a mass term to the 2d chiral multiplets
\cite{Hanany:1997vm}.

Now, let us turn on a parameter that corresponds to moving the
NS5$'$-brane (and, therefore, the D2-brane) in the $x^6$
direction. It forces the D2-brane to end on one of the
D4-branes, {\it cf.} Figure \ref{dbranefig} $a$. From the point
of view of the 2d theory on the D2-brane it
corresponds to turning on the Fayet-Iliopoulos parameter of the
$U(1)$ gauge group \cite{Hanany:1997vm}. Depending on the sign
of the FI term, either the chiral fields of charge $+1$ or of
charge $-1$ gain expectation values, connecting the D2-brane
and either set of the D4-branes. To match the brane picture, it
must be the case that the cubic superpotential coupling will
insures that only one of the two types of fields can receive
expectation values. Indeed an expectation value for both types
of fields would act as a delta-function source for the four
dimensional hypermultiplet fields. In the Coulomb branch of the
theory, expectation values for the Higgs branch fields, which
are massive, will typically break SUSY.

If all the parameters $a_i$ are set to zero,
the space of vacua in such a theory is the K\"ahler quotient
$$
\C^N // U(1) \cong \cp^{N-1} \,.
$$
The K\"ahler modulus can be combined with a B-field $\eta$ on
the target space $\cp^{N-1}$ to a complexified FI parameter $t$.
In $\CN=(2,2)$ 2d theories, such as the one we are
considering, the values of the Fayet-Iliopoulos parameter
$\alpha$ and the theta-angle $\eta$ are renormalized due to
quantum corrections. The renormalized value of the complex
parameter $t$ can be expressed in terms of the twisted superpotential:
\begin{equation}\label{fiviaw}
t = {\p \CW (\sigma) \over \p \sigma} \end{equation}

\begin{figure}[b] \centering \includegraphics[width=2.0in]{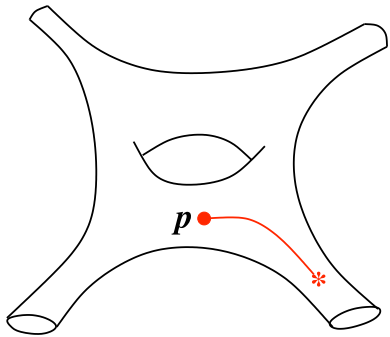} \caption{The effective twisted superpotential $\CW (v)$ of the D2-brane
theory can be expressed as an integral over an open path on the SW curve.\label{intfig}} \end{figure}

As was explained above,
in the brane construction the classical (``bare'') FI parameter $\alpha$
is identified with the position of the D2-brane in the $x^6$ direction,
{\it cf.} Figure~\ref{branefig} $a$,
while its ``quantum'' companion $\eta$ can be
identified with the position in the $x^{10}$ direction
(which is not manifest in the type IIA theory/classical field theory).
As in \cite{Witten:1997sc}, we can describe quantum corrections by
performing the usual M-theory lift of this picture and identifying
the effective value of the complexified FI parameter $t$ in the IR theory
with the distance between the M2-brane and the M5-brane
in the complex plane parameterized by $x^{10} + i x^6$,
\begin{equation}\label{fiviaxx} t = \Delta x^{10} + i \Delta x^6 \,. \end{equation}
Combining this with eq. (\ref{fiviaw}), the identification
of the position in the $v$ plane with the twisted chiral field $\sigma$,
and the fact that the M5-brane worldvolume is the SW curve $\Sigma$,
we arrive at the following property of the twisted superpotential:
\begin{equation}\label{wintsw}\frac{\partial \CW_{D2} (a,v)}{\partial v} = t \,. \end{equation}
This expression is actually equivalent to (\ref{eq:super}).
Indeed, eq. (\ref{eq:super}) represents the effective superpotential
for the bulk fields after integrating out $\sigma=v$.
By the standard rules of the Legendre transformation,
the solution to (\ref{wintsw})
is the derivative of the effective superpotential $\CW$
with respect to $t=z$, {\it i.e.} $\lambda$.

The linear sigma model construction we meet here has some
interesting features, and an unpleasant one.
On the one hand, it gives a slightly better definition of surface operators
than the one based on a codimension two singularity for the gauge field,
or the coupling of the 4d
$SU(N)$ gauge theory to a $\cp^{N-1}$ sigma model on $\cS$.
The advantage of
the linear sigma model is that it allows one to follow
the ``flop'' from positive to negative values of $\alpha$,
which appears to relate surface operators for consecutive
gauge groups in the quiver.
On the other hand, it is still not as powerful as one might desire:
{\it e.g.} in the $N=2$ case,
where the $U(1)$ flavor symmetry of the bifundamental field
is promoted to a crucial $SU(2)$ flavor group,
one has to gauge this symmetry group to produce a generalized quiver.
The cubic superpotential coupling of the bifundamental field
(in the $N=2$ example, an $SU(2) \times SU(2) \times SU(2)$ ``trifundamental'')
to the 2d chiral multiplets
cannot preserve this extra $SU(2)$ symmetry.

It would be interesting to find a description of the surface
operator capable to describe in a symmetric fashion all three
possible ``flops'' which may transport the basic surface
operators of either of the three $SU(2)$ groups through the
pair of pants.

We have now a rough, self-consistent picture of the
correspondence between six and four dimensional surface
operators, in a given weakly coupled four dimensional
Lagrangian description. Well inside a tube the surface operator
should be well described by the basic defect operator where the
$SU(2)$ gauge group corresponding to the tube is broken to a
$U(1)$ subgroup. Near the endpoints of the tube, the pure gauge
theory description breaks down, and the defect is better
described by coupling to a 2d sigma model, associated to a
specific pair of pants. Flops in the 2d sigma model connect the
surface operators living on different legs of the pants. We
will not attempt to refine this picture further in this paper.

\subsection{Instanton counting}

Now we are ready to discuss instanton counting in the presence of
a surface operator.
In particular, our goal is to clarify
the claim, made in the previous sections, that the semiclassical
behavior of the Nekrasov partition function in the presence of
a surface operator\footnote{It is worth noting that in \cite{Braverman:2004vv}
it has been independently proposed that the instanton
partition function in the presence of a surface operator should
satisfy a differential equation of the type (\ref{degendiff}).
It would be interesting to explore further the connections with that work.}
matches the semiclassical limit
of the conformal block with the insertion of a degenerate field,
and to set the stage for a computation beyond the semiclassical limit
(that we will not attempt in the present paper).

Following \cite{Nekrasov:2002qd}, we introduce the generating function
\begin{equation}
\label{zinst}
\CZ_{{\rm inst}} (a,q,\epsilon;\LL,t)
= \sum_{k=0}^{\infty} \sum_{\m \in \Lambda_{\LL}} q^k e^{i t \cdot \m}
\oint_{\CM_{k,\m}} 1
\end{equation}
where $q = e^{2 \pi i \tau}$
and $\CM_{k,\m}$ is the moduli space of ``ramified instantons.''
{}From the point of view of the 4d gauge theory
(where a surface operator supported on $\cS$ is defined as in (\ref{asing}) and  (\ref{etaphase}))
the ramified instantons are anti-self-dual gauge connections on $\IR^4 \setminus \cS$
with instanton number $k \in \Z$ and monopole number~$\m \in \Lambda_{\LL}$.

As noted above, one can also represent surface operators of Levi type $\LL$
by studying 4d gauge theory on $\IR^4$
coupled to a 2d sigma model
on $\cS \subset \IR^4$ with the target space $G / \LL$.
In this description, %which will be especially useful to us in instanton counting,
the complex parameter $t$ is the complexified K\"ahler
modulus of the flag manifold $G / \LL$ and ``ramified instantons''
with $\m \ne 0$ can be thought of as the usual instantons
of the 4d gauge theory combined
with 2d worldsheet instantons of the sigma model.
Indeed, according to (\ref{mlattice}), the monopole number
$\m \in \Lambda_{\LL} \cong H_2 (G/\LL ; \Z)$
measures the degree of the map $\Phi: \cS \to G / \LL$.
In the case we are mostly interested in,
where $\LL$ is the next-to-maximal Levi subgroup,
we have $G / \LL \cong \cp^{N-1}$ and
the monopole number is simply an integer, $\m \in \Z$.

The moduli space $\CM_{k,\m}$ is non-compact, so the integral
in (\ref{zinst}) needs to be properly defined (regularized).
This can be achieved by noting that $\CM_{k,\m}$ admits a
natural action of the gauge group $G$ (which acts by a change
of framing at infinity) and an action of the 2d
torus ${\bf T}^2_E$ (induced by the action of ${\bf T}^2_E =
SO(2)_1 \times SO(2)_2 \subset SO(4)$ on $\IR^4$). Therefore,
the integral on the right-hand side of (\ref{zinst}) can be
conveniently regularized by considering the equivariant
integral of the unit $G \times {\bf T}^2_E$-cohomology class
over $\CM_{k,\m}$. This integral takes values in the field of
fractions of the ring $H^*_{G \times {\bf T}^2_E} ( {\rm pt}
)$, which can be identified with the ring of functions on the
Cartan subalgebra of $G \times {\bf T}^2_E$, invariant under
the Weyl group. Therefore, the equivariant integrals on the
right-hand side of (\ref{zinst}) are rational functions of $a$,
$\epsilon_{1}$, and $\epsilon_{2}$, where $a = (a_1, \ldots,
a_N)$ and $\epsilon_{1,2}$ denote coordinates on the Lie
algebra of $\TT \subset G$ and ${\bf T}^2_E$, respectively.

As in \cite{Nekrasov:2002qd}, combining the instanton partition
function with the classical term and the one-loop term we
obtain the full partition function,
\begin{equation}
\label{zfull}
\CZ_{4d} = \CZ_{{\rm classical}} \cdot \CZ_{{\rm 1-loop}} \cdot \CZ_{{\rm inst}}
\end{equation}
that we already encountered in Section 2. As we claimed there,
the general structure of conformal blocks with degenerate field
insertions match the semiclassical expansion of the partition
function $\CZ_{4d}$ in the presence of surface operators,
\begin{equation}
\CZ_{4d} \, \sim \, \exp \Bigl(\, -\frac{\CF(a_i)}{\epsilon_1 \epsilon_2}
+ \frac{\CW(a_i,t)}{\epsilon_1} + \ldots \, \Bigr)\, ,
\end{equation}
where the prepotential $\CF(a_i)$ and the twisted superpotential $\CW(a_i,t)$
are the F-terms of the 4d theory on $\IR^4$
and the 2d theory on~$\cS = \IR^2$ that contribute to the Nekrasov partition function.
Indeed, by the localization rule
$$
{\rm Vol} (\IR^4) = \int_{\IR^4} 1 = \frac{1}{\epsilon_1 \epsilon_2}
\qquad , \qquad
{\rm Vol} (\IR^2) = \int_{\IR^2} 1 = \frac{1}{\epsilon_1}
$$
where we assumed that the surface operator is supported on
a plane~$\cS = \{ w_1 = 0 \}$. For a surface operator supported at $w_2=0$
the roles of $\epsilon_1$ and $\epsilon_2$ are exchanged.
As we explained in Section 2, in the Liouville theory these
surface operators correspond to the degenerate fields $\Phi_{2,1}$ and $\Phi_{1,2}$.

Notice that the surface operator breaks the permutation
symmetry between the $(a_1, \ldots, a_N)$. In particular, the
classical twisted superpotential will be written as $(\eta +
\tau \alpha) a_i$ for a certain choice of $i$. More generally,
the instanton partition function is not invariant under Weyl
group permuting the $(a_1, \ldots, a_N)$, unless one acts on
this extra dummy label $i$ as well. This is as it should be to
match the conformal block interpretation. Conformal blocks
without a degenerate insertion are labelled by continuous
(Liouville or Toda) momenta, each subject to the identification
by the action of the Weyl group ($\alpha \to Q-\alpha$ for
Liouville theory). Conformal blocks with a degenerate insertion
in a certain leg carry an extra discrete label: the momenta on
the two sides of the degenerate insertion must differ by a
value allowed by the degenerate fusion rule. The Weyl group
acts non-trivially on this difference. This discrete label
coincides with the extra dummy label in the instanton partition
function.

More generally, we believe that,
for every conformal block with a degenerate field insertion,
there should be a half-BPS surface operator supported on
a surface $\cS \subset \IR^4$ invariant under the symmetry (\ref{torusact}).
The definition of such surface operator should be given
in the corresponding generalized $SU(2)$ quiver gauge theory,
and allow for a computation of the Nekrasov partition function
in the presence of the surface operator.
In particular, it is natural to expect that the degenerate field $\Phi_{2,2}$
corresponds to a surface operator supported on a degenerate curve $\cS$
defined by the equation $w_1 w_2 = 0$.

\section{Line operators on surface operators}\label{sec:line-on-surface}

$\CN = (2,2)$ theories in two dimensions have interesting
half-BPS line operators. They preserve the diagonal combination of
$SU(1,1|1)_L \times SU(1,1|1)_R$. A useful way to produce such
line operators is to consider a deformation of the theory where
some marginal coupling $t$ has a non-constant profile $t(x^1)$
as a function of the space coordinate $x^1$ over a finite
region $-L<x^1<L$. A flow to the IR sends the scale $L \to 0$
and squeezes the profile $t(x^1)$ to a step function.

The resulting line operators are labeled by  the path in the
space of couplings, up to homotopy. This construction applies
as well to the construction of line operators inside surface
operators. A simple, rich example appears in
\cite{Gukov:2006jk} in the case of $\CN=4$ super Yang-Mills. We are
especially interested in line operators for which the path in
the space of couplings is closed, so that the line operator
does not interpolate between two distinct surface operators.

It is easy to understand the meaning of such line operators in
an Abelian gauge theory. If we consider a profile for the
coupling $\eta$ to the magnetic flux, and we write
$\eta(x^1)=\eta_0+ \delta \eta(x^1)$ with $\delta \eta(\pm
\infty)=0$, we get a term in the Lagrangian
\begin{equation}
\int \eta(x^1) F = \int \eta_0 F - \int d \delta \eta \wedge A
\end{equation}
In the IR the latter term reduces to $\Delta \eta \int dx^0
A_0$. (We take the surface operator to span $x^0,x^1$). This
line operator coincides with the insertion of a Wilson line for
the $U(1)$ gauge group! A similar reasoning (or a simple electromagnetic
duality) shows that a discontinuity $\Delta \alpha$ coincides
with the insertion of a 't Hooft line operator.

We can use this result in two ways. In a non-Abelian gauge
theory where the surface operator breaks the gauge group to,
say, $\LL = SU(N-1) \times U(1)$, the Wilson and 't Hooft line
operators will live in the $U(1)$ factor. These operators are
defined independently from the bulk line operators. However we
will learn how to reproduce the bulk line operators from line
operators living on a surface operator.

In the Coulomb branch of the non-abelian gauge theory, the line
operators will take the form of 't Hooft-Wilson line operators
with charges $q_i=\Delta \eta_i, p^i=\Delta \alpha^i$. As the
parameter space of surface operators in the IR coincides with
the SW curve $\Sigma$ , one could consider line
operators in the IR associated to a closed path $\gamma$ on
$\Sigma$, which carry charge
\begin{equation}
q_i + \tau_{ij} p^j = \oint_\gamma \omega_i.
\end{equation}
Alternatively, $\gamma = q_i \alpha^i + p^j \beta_j$ in a
canonical basis of one-cycles.

From the six-dimensional point of view, as our surface
operators are labeled by a point in the curve $C$ over which
the twisted $(2,0)$ theory lives, we expect to see line
operators labeled by closed paths in $C$, up to homotopy.
We have two rather distinct ways to label a line operator attached
to a surface operator: a homotopy class of paths in $C$ in the
UV 6-dimensional theory, and a homology class in $\Sigma$ in
the IR theory. We already encountered this phenomenon in
Liouville theory, and understood the relation to the WKB
analysis of \cite{Gaiotto:2009hg}: expectation values of UV
line operators are linear combinations of individual
contributions, each taking the form expected from an IR line operator.

Now we are ready to provide explicit expressions for the 2d CFT
operators which represent the action of line operators on surface operators.

\subsection{Line operators from braiding and fusion}
In order to introduce line operators in a setup where
localization is possible, we need the support of the line
operator to be invariant under the two relevant $U(1)$
isometries. The isometries are the rotation in the plane of the
surface operator, and the rotation in the plane orthogonal to
the surface operator. Although until now we mostly referred to
straight line operators, a conformal transformation allows us
to consider circular line operators as well.
In complex coordinates $z,w$ on $\C^2 \cong \IR^4$, we can consider a
surface operator at $w_2=0$ with a line operator at $|w_1|=1$.
The same location works for $S^4$, in stereographic
coordinates.

Given a conformal block with the insertion of a degenerate
field $\Phi_{2,1}(z)$, we can ask: what is the effect of
transporting the point $z$ along a closed path $\gamma$ on the surface
$C$? This is a well studied problem in the context of rational
conformal field theories \cite{Moore:1989vd}.
If we insert the operator
$\Phi_{(2,1)}$ in a certain channel of the conformal block, the
result is (by definition) a power expansion in $z$, which is
convergent as long as $z$ lies in the corresponding tube of the
Riemann surface. The conformal block is defined outside that
region by analytic continuation. The analytic continuation is
naturally executed stepwise, by moving $z$ from a tube to an
adjacent one. Such elementary moves are represented by $2
\times 2$ matrices acting on the corresponding spaces of
conformal blocks. We refer to Appendix
\ref{app:fusionrules} % and \ref{app:degfusion}
for a discussion of this fact, and a
review of the explicit calculation of the fusion and braiding
matrices.

In order to understand the elementary moves, we just need to
consider the simplest possible setup, where a single degenerate
insertion moves between the three legs of a three-point vertex
of full punctures. This has the physical interpretation of a
surface operator in the ``pair of pants'' theory of four free
hypermultiplets, with masses turned on in the Cartan of the
$SU(2)_1 \times SU(2)_2 \times SU(2)_3$ flavor subgroup.

If we place the full punctures of Liouville momenta
$\alpha_1,\alpha_3,\alpha_4$ respectively at $0$,$1$, $\infty$
on the sphere, and the degenerate insertion at $z$, the
conformal blocks can be given explicitly in terms of
hypergeometric functions. The basis of conformal blocks where
the degenerate field is inserted, say, on the $a_1$ leg behave
as $z^{\Delta_{\alpha_1 \pm b/2} - \Delta_{\alpha_1} -
\Delta_{(2,1)}} = z^{b (\frac{Q}{2} \mp \alpha_1)}$ as $z \to
0$. The transformation of basis to solutions with well defined
behavior near $z=1$ is called fusion matrix, and will be
denoted as $F_{\pm \pm}$. This has to be intended as a
transport along the positive real axis. The transformation of
basis to solutions with good behavior as $z\to \infty$ is
called braiding matrix, and will be denoted as $B_{\pm \pm}(\pm
1)$. The sign $\pm 1$ refers to transport from $0$ to $\infty$
along the positive real axis, on either side of $z=1$.

For more general conformal blocks, we need to set up a useful
convention to distinguish the continuous labels at the
intermediate channels from the discrete $\pm$ label associated
to the $\pm \frac{b}{2}$ shift. In order to do that, we add a
dummy label to the conformal block: we do not just specify in
which leg of the conformal block we insert the degenerate
field, but also ``near which end'' of the leg. Thus we label
the conformal block by the Liouville momentum $a$ through the
``long'' piece of the leg. The other, ``short'' part of the leg
has momentum $a \pm \frac{b}{2}$. When a degenerate insertion
is moved from one end to the other of the same leg, the notions
of ``long'' and ``short'' parts of the leg are exchanged, and
the continuous label is shifted by $\pm \frac{b}{2}$.

The transport of the degenerate insertion along a path in the
Riemann surface gives a sequence of elementary operations:

\addtolength{\baselineskip}{-1mm}
\begin{itemize}
\item fusion and braiding matrices, which only act on the
    discrete label
\item transport along a leg, which act by a diagonal shift
    operator $a_i \to a_i \pm \frac{b}{2}$
\item transport around a leg, which provides a diagonal
    phase factor.
\end{itemize}

\addtolength{\baselineskip}{1mm}

It is rather simple to connect this decomposition to the
semiclassical approximation in the perturbative regime. In that
regime, the branch points of the cover $\Sigma \to C$ lie in
the pair of pants regions, away from the long, thin tubes
associated to the $SU(2)$ gauge groups. It is easy to see from
the expression of $\phi_2$ for the pair of pants theory that
each pair of pants in $C$ supports a single cut in the branched
cover $\Sigma \to C$. Only when the path $\gamma$ in $C$ passes
through a pair of pants there is some ambiguity on the lift
$\tilde \gamma$ in $\Sigma$.  The $2\times 2$ fusion and
braiding matrices differentiate between the two possible
choices of sheet of $\Sigma$ entering and exiting the pair of
pants. The transport along and around the tube is perfectly
diagonal, and well described by the naive WKB analysis.

\subsection{S-duality of line operators in $N_f=4$ theory}

As an illustrative application of the Liouville CFT technology, let
us consider the loop operators, acting on a surface operator, in
$N_f=4$ $SU(2)$ gauge theory, for which the instanton partition
function coincides with the Liouville conformal blocks of the four punctured sphere.
As usual, we place the punctures, of momenta $\alpha_1$,$\alpha_2$,$\alpha_3$,$\alpha_4$ respectively at
$0,q,1,\infty$, and consider a trivalent graph connecting $0,q$ and $1,\infty$ by a channel of momentum
$\alpha$. We will now introduce the Wilson and 't Hooft loop operators via the CFT monodromy operation,
and explicitly demonstrate that S-duality interchanges the two.

The surface operator is represented by  a (2,1) degenerate operator placed
at some location $z$ on one of the legs of the conformal block. For definiteness,
we place it on the internal leg of the conformal block, say, near
the $1,\infty$ vertex.  As will become clear below, this choice is particularly
convenient for studying the action of S-duality.

\begin{figure}[t] \centering \includegraphics[width=5.7in]{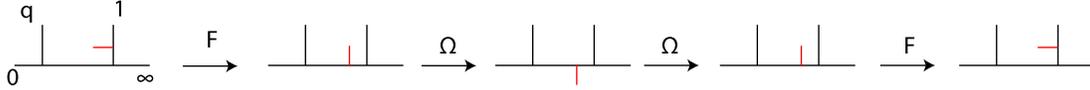} \caption{The Wilson loop move. The red line represents the degenerate insertion} \label{wilsonl}\end{figure}

The basic Wilson line operator transports the degenerate field around the internal leg.
In the notation of the appendix \ref{app:fusionrules}, this produces a simple phase factor \begin{equation}\left( \Omega_{\alpha,-\frac{b}{2}}^{\alpha \pm \frac{b}{2}}\right)^2 = e^{ 2 \pi i b ( \mp Q /2 - Q /2 \pm  \alpha)} = e^{2 \pi i b ( - Q /2 \pm  \frac{a}{\hbar})} \end{equation}
To do S-duality, we need to apply a fusion matrix ${\cal F}$ that maps the original `s-channel' conformal block
into a `t-channel' block, associated to a graph where $(0,\infty)$ and $(q,1)$ are joined by an internal leg of momentum $a'$.
If, during this operation, we want to keep the degenerate field insertion in the intermediate channel, we need to specify
in detail the relative motion of the punctures at $z$ and $q$. It is simpler to
move $z$ away from the intermediate leg, and place it, say, on the $q=1$ external leg.
With this choice,  the Wilson line operator takes the schematic form
\be
W=F \, \Omega^2 F,
\ee the degenerate insertion is transported (via a fusion matrix $F$) to the intermediate leg,
rotated around it via $\Omega^2$, and then fused back to the external leg (see Fig. \ref{wilsonl}).

\begin{figure}[b] \centering \includegraphics[width=3.4in]{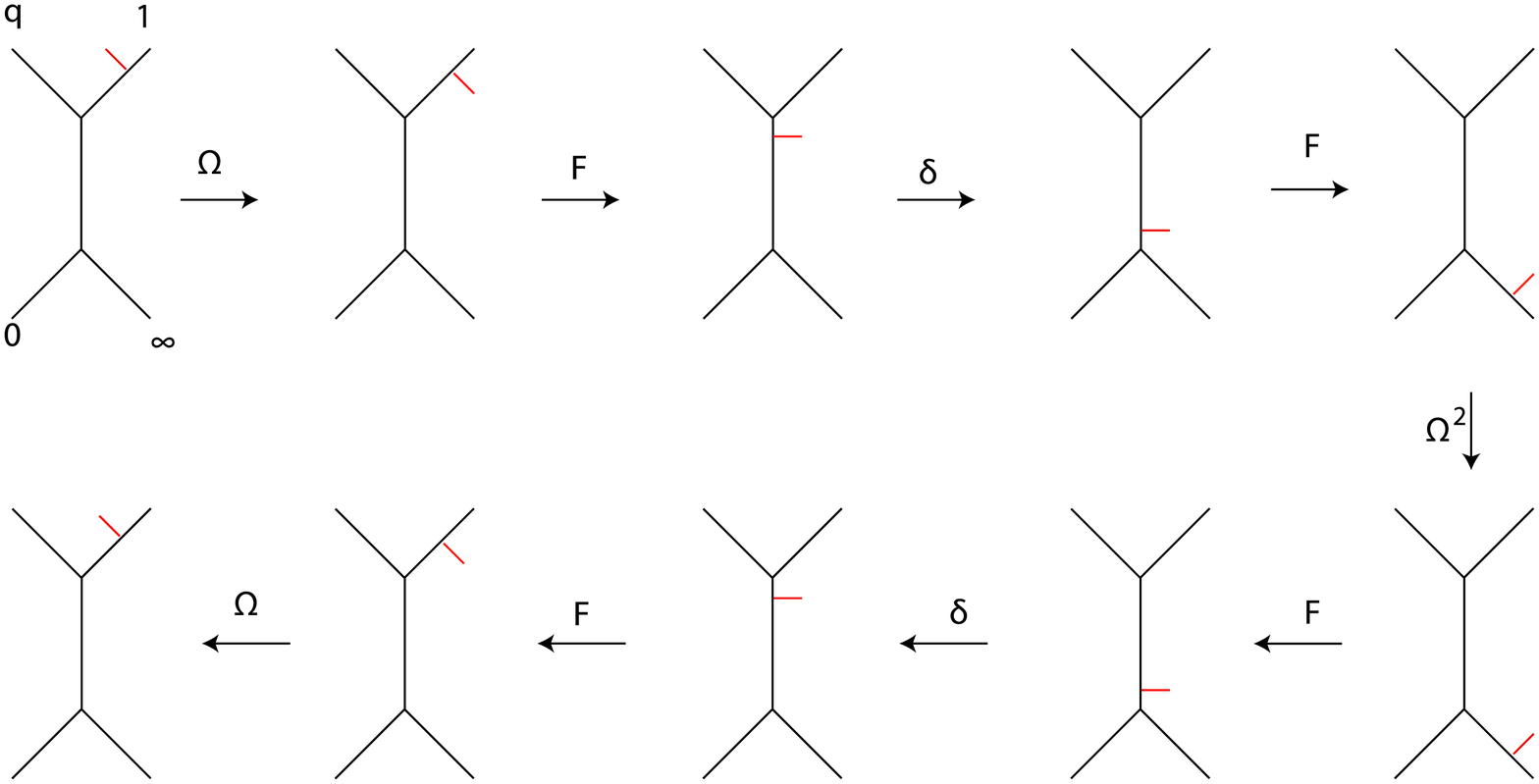} \caption{The 't Hooft loop move. It represents the same path as in the previous picture, but in the S-dual frame} \label{figthooftl}\end{figure}

A priori, we could now compute the 't Hooft loop expectation values by defining them as
the Wilson loops in the S-dual theory, and by using the known form of the fusion matrix ${\cal F}$ that implements
the S-duality transformation on the Liouville conformal blocks:
\be
\label{Ffusion}
\inctt{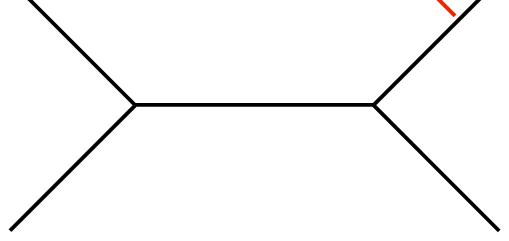} \ \ = \ \  {\cal F} \ \ %\sum_{a'} F_{aa'}\mat{a_2}{a_3}{a_1}{a_4}
\inctt{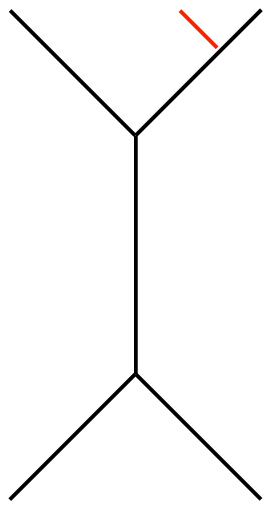}\, .
\ee
In other words,  the 't Hooft loop $H$ could be obtained by commuting the Wilson loop $W$ with ${\cal F}$
\be
\label{wfisfh}
W {\cal F} =
{\cal F} H
\ee
However,  the fusion matrix ${\cal F}$ for arbitrary conformal blocks is quite involved, and
this type of calculation is hard to do in practice.
So instead, we
define the 't Hooft loop $H$ more directly, via the monodromy operation of a degenerate field on the S-dual path,
as indicated in fig \ref{figthooftl}.
Schematically, the sequence of moves that defines $H$ is (c.f. fig. \ref{figthooftl}):
\be
\label{hmon}
H= \Omega F \delta F \Omega^2 F \delta F \Omega,
\ee
 the degenerate insertion is rotated to the other side of $1$, fused to the internal leg,
transported across it, rotated around $\infty$, transported back, fused back to the $1$ external leg, and rotated to the original configuration. Because of the two shift operators, the final expression contains three different terms, where $a'$ is subject to shifts $\pm b,0$.

\begin{figure}[t] \centering \includegraphics[width=5.9in]{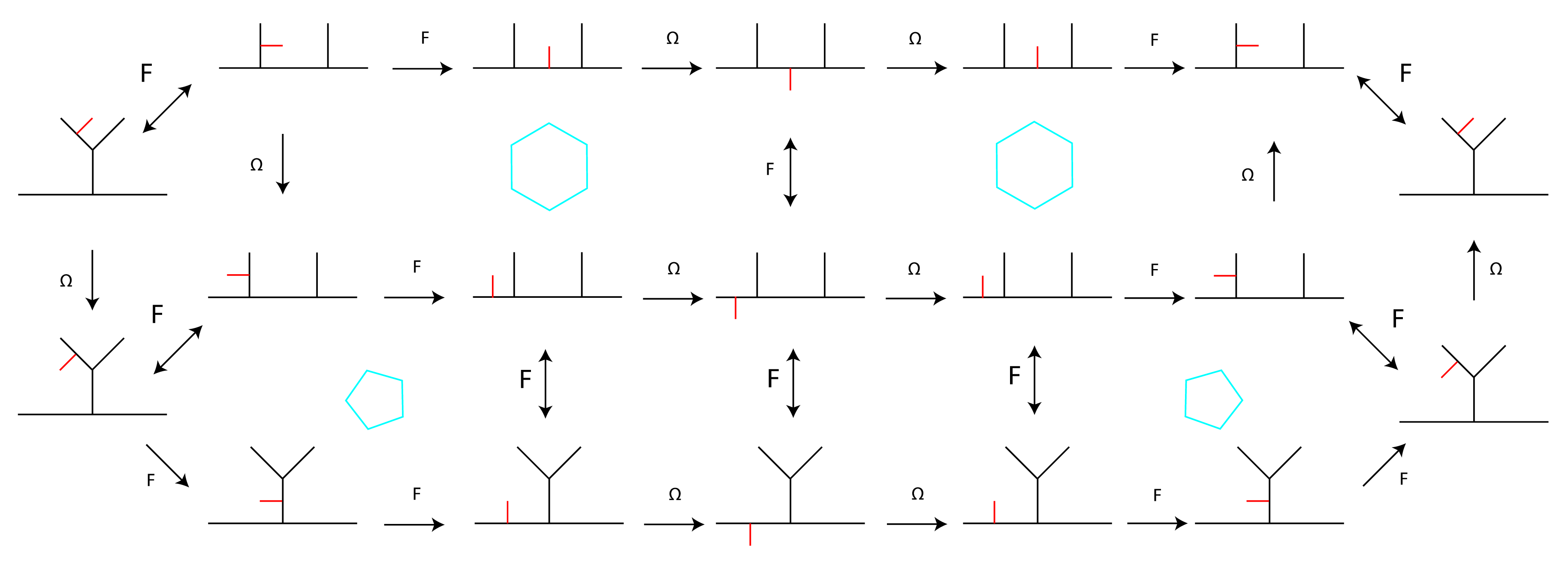} \caption{The Moore-Seiberg transformations which verify S-duality for the line operators.
The upper row gives  the computation of a Wilson loop in $N_f=4$. The computation is manipulated through hexagon and pentagon relations. 
 \label{sduality}} \end{figure}

The monodromy operation in Fig. \ref{figthooftl} involves relatively simple
braiding and fusion matrices, that do not act on the modular parameter $q$, that defines
the $SU(2)$ gauge coupling. Moreover, the braiding and fusion matrices can be shown to satisfy important consistency relations, known as the pentagon and hexagon identities, which among others can be used to derive the S-duality
relation (\ref{wfisfh}). The relation is proved graphically in Fig.~\ref{sduality}. 

Here we sketch the algebraic steps. % as depicted in fig (\ref{sduality}).
First, we expand:
${\cal F} H = {\cal F} \Omega F \delta F \Omega^2 F \delta F
\Omega= \Omega ({\cal F} F \delta F) \Omega^2 F \delta F
\Omega$. We apply the pentagon identity to the block in
parenthesis ${\cal F} H = \Omega (F {\cal F} ) \Omega^2 F
\delta F \Omega$ and commute ${\cal F}$ through ${\cal F} H =
\Omega F \Omega^2 ({\cal F} F \delta F) \Omega$. Another
pentagon identity and commutation brings us close to the final
result ${\cal F} H = (\Omega F \Omega)(\Omega F \Omega) {\cal
F}$. Finally, two hexagon relations give ${\cal F} H =  F \Omega F^2
\Omega F  {\cal F} = F \Omega^2 F {\cal F} = W {\cal F}$.

\section{Line operators}\label{sec:line}

We are now ready to study the gauge theory line operators that act in the bulk, without any (nearby) surface operators.
Such bulk line operators a priori look quite different from the line operators that act on a surface operators.
Surface line operators are essentially abelian, since (for a surface operator with a next-to-maximal Levi subgroup)
they live in a single $U(1)$ factor of the gauge group $G$,
%$SU(N-1)\times U(1)$,
whereas bulk line operators are non-abelian.
Nonetheless, we claim that a bulk line operator can be obtained by annihilating two identical surface operators,
 one of which contains a surface line operator.\footnote{To see this, recall that a surface operator
restricts the gauge transformations to the subgroup that, at the surface, commutes with
the $U(1)$.
Annihilating two surface operators reinstates the full  gauge symmetry.
The bulk loop %$W_B$
is given by averaging the surface loop % $W_S$
over the full gauge orbit.
Via standard coadjoint orbit quantization, this yields a non-abelian loop operator.
}
In the Introduction, we used this insight, combined with the 6d perspective, to argue that the Wilson-'t Hooft
loops in our class of gauge theories can be identified with certain loop operators in Liouville CFT,
defined in terms of the four step monodromy procedure summarized in Section 1.2.
In this section, we will use this identification to compute the expectation values of Wilson and 't Hooft loops for
certain basic examples. First, we will present an independent motivation for our proposed CFT definition
of the line operators.

\subsection{Wilson-'t Hooft loops from Liouville CFT}

Consider
a circular Wilson line  $W^{(k)}_{j}$ in the spin $j$ representation of the $k$-th $SU(2)$ gauge group factor.
As shown by Pestun, inserting $W^{(k)}_{j}$ inside the gauge theory instanton sum on $\IR^4$, i.e. with
given Coulomb branch parameters $a_i$, simply amounts to
multiplication by the corresponding character
\be
\label{wilsonc}
W_{j}(a_k) = {\rm tr}_{R_j}\bigl( e^{4\pi i b a_k T_3 }\bigr)\, = \sum_{p=-j}^{j}e^{2 \pi b p a_k}.
\ee Here and in the following, the summation $\sum_{p=-j}^j$ for the half-integral $j$
stands for the sum over half-integral $p$ between $-j$ and $j$.
Instanton sums on $\IR^4$ are therefore eigenfunctions of the circular Wilson line operators.
On $S^4$, the Wilson loop expectation value takes the form
\be
\label{wilsont}
\bigl\langle \, W_j^{(k)} \, \bigl\rangle_{{}_{\rm \! \! S^4}} = \int \!  da_i \, \bigl|\PPsi(\tau; a_i ) \bigr|^2  W_j(a_k)
\ee
A similar direct gauge theory calculation of the expectation value of 't Hooft loop operators in $\CN=2$ gauge theories
is not yet available. However, based on the semi-classical discussion of section 2, we expect that these will take the following
schematic form
\be
\bigl\langle \, H_j \, \bigl\rangle_{{}_{\rm \! \! S^4}} = \int \! da_i da'_k\, \overline{\PPsi(\tau; a_i)} \PPsi(\tau; a_k') H_j(a_i,a'_k) ,
\ee
where $H_j$ denotes some 't Hooft loop associated with the spin $j$ representation. % in the $k$-th $SU(2)$ factor.
In the following we will explicitly compute the kernel $H_{j}(a,a')$ in some specific examples.

The Wilson and 't Hooft loops are special cases of a more general class of dyonic 't Hooft-Wilson line operators, whose systematic study was initiated in \cite{Kapustin:2005py,Kapustin:2006hi}.
We can make contact here with the recent work
\cite{Drukker:2009tz}, which provides a useful classification
of 't Hooft-Wilson line operators in generalized $SU(2)$ quiver
gauge theories. The operators are labeled by a set of magnetic and electric charges $p_i$ and $q_i$
for each $SU(2)$ gauge group,  subject to an identification $(p_i,q_i) \to (-p_i, -q_i)$ for each $i$,
and to a constraint: the sum of the three magnetic charges $p_i$ for the three $SU(2)$ gauging a single matter block should be even.\footnote{The authors of \cite{Drukker:2009tz} find it useful to enlarge the space of line operators, by including
magnetic flavor line operators. Here, for now, we only consider
line operators for the gauge groups.} The authors of \cite{Drukker:2009tz} propose a suggestive identification between the
set of line operator charges and the set of (homotopy classes of) closed non-selfintersecting
curves in C. %The action of S-duality over such space is described in detail.
This classification via closed non-selfintersecting paths $\gamma$ on $C$
naturally fits with
the description of the loop operators via the  6-d (2,0) theory
as the end-point of supersymmetric
semi-infinite M2-branes, as reviewed in the Introduction.

Correspondingly, we will denote a general Wilson-'t Hooft loop operator %associated with a path $\gamma$
by
\be
\Phi_j(\gamma)\, .
\ee
Here the label $j$ indicates the spin $j$ of the $SU(2)$ representation.
In the gauge theory, the operators $\Phi_j(\gamma)$ can be thought of as the effect of transporting a dyonic
point particle, with charge labeled by the path $\gamma$, around the loop trajectory. 

 In a given perturbative
regime, one can identify a complete set of non-intersecting cycles on $C$, that lift to a complete set of
$A$-cycles on the SW surface $\Sigma$. We will denote this set of cycles by $A_k$.
On the SW  surface, we can choose a set of dual $B$-cycles, that project back to $C$ to a set of
dual cycles that we denote by $B_k$.\footnote{This description is slightly oversimplified. The $A$ and $B$-cycles lie on the Prym curve of $\Sigma$. Moreover, the set of B-cycles is not unique. E.g. one is free to
apply shifts of the form $B_k \to B_k+A_k$. In the gauge theory, this freedom is a reflection of the Witten effect,
shifts in the dyonic charge spectrum by one electric unit. For small theta angles $\theta_k$, however, there is a
preferred choice of dual B-cycles that correspond to the pure monopole charges.}
The Wilson and 't Hooft loops are then identified as the loop operators
associated with the A and B-cycles, respectively
\be
W_j^{(k)}  \equiv \Phi_{j}({A_k})\, , \qquad \qquad
H_j^{(k)}  \equiv \Phi_{j} (B_k)\, .
\ee

If one would consider
the insertion of multiple Wilson lines (all located on concentric circles, invariant under the  $U(1)$ rotation that is used
to justify the localization of the gauge theory path integral),
one would discover that the Wilson lines form a commutative and associative algebra, given by the representation
ring of $SU(2)$:
\be
W_{\ell}^{(k)}(a)  \, W_{s}^{(k)}(a) \, = \, \sum_{|\ell-s| \leq\, j\, \leq |\ell +s|} W_{j}^{(k)}(a).
\ee
Via S-duality, we thus learn that, in general, all operators $\Phi_j(\gamma)$ associated with some given path $\gamma$ form a commutative associative algebra, isomorphic to the $SU(2)$ representation ring. 
More generally, we will see that two line operators $\Phi_{j_1}(\gamma_1)$ and $\Phi_{j_2}(\gamma_2)$
do not commute in case the two curves $\gamma_1$ and $\gamma_2$ intersect.

We now set to
describe the identification between line operators of charge $\gamma$ and the Verlinde loop operators associated with the same path $\gamma$ on $C$.\footnote{In fact, we face a small puzzle here: as we will see shortly, the loop operators in the CFT make sense for all $\gamma$,
while the gauge theory line operators appear to require the loop $\gamma$ to be non-self-intersecting.
We will propose a resolution to this puzzle in Appendix D.}
Within the context of rational CFTs, the Verlinde operators are known to generate
a commutative and associative algebra, given by the fusion algebra of the CFT. Here we would like
to define analogous operators in Liouville theory.  

Liouville CFT is a non-rational CFT. It has a continuous spectrum of primary operators. Furthermore, part of the 
operator spectrum, including the identity operator, create non-normalizable states. It is not at all obvious, therefore, 
that Liouville theory possesses a well-defined fusion algebra, with similar properties to that of a rational CFT. 
However,  the discrete sub-spectrum of degenerate Virasoro representations
do seem to specify a well-defined closed sub-algebra.
In particular, the Virasoro modules associated with the operators $\Phi_{n,1}$ generate a closed fusion algebra
\be
[\, \Phi_{p,1}] \, \times \, [ \, \Phi_{q,1} \, ]\, = \, \sum_{|p-q+1| \leq\, n \, \leq |p+q-1|} [ \, \Phi_{n,1} ] .
\ee
This algebra is identical to the representation ring of $SU(2)$, via the identification
$n \equiv 2j+1\, .$
 More generally, the fusion algebra of a degenerate field
with a continuous representation also seems well defined. It reads (here $j \equiv \frac{n-1}{2}$)
\be
[\, \Phi_{n,1}] \, \times \, [ \, V_a \, ]\, = \,\sum_{p=-j}^j
 [ \, V_{a+  p b} ]
\ee
where $[V_a]$ denote the chiral sub-Hilbert space with given Liouville momentum $a$.
Here and in the following, the symbol $\sum_{p=-j}^j$ for (half-)integral $j$
stands for the summation over $p=-j, -j+1, \ldots, j-1, j$ as usual.

\smallskip

\noindent
We now define the Verlinde monodromy operators $\Phi_{j}(\gamma)$  via the following 
recipe~\cite{Verlinde:1988sn}:

\vspace{-2.5mm}

\begin{enumerate}
\item Insert the identity operator $\mathbf{1}$ inside a chiral  Liouville correlation function.
\vspace{-2mm}
\item Write  $\mathbf{1}$  as the result of fusing two chiral operators $\Phi_{2j+1,1}(z)$,
via their OPE.
\vspace{-2mm}
\item Transport one of the operators along a closed non-self-intersecting path $\gamma$.
\vspace{-2mm}
\item Re-fuse  the two degenerate fields together into identity $\mathbf{1}$, via their OPE.
\end{enumerate}

\vspace{-2.5mm}

\noindent
This procedure defines a linear map on the space of Liouville conformal blocks.
We need to introduce a normalization factor $\cN_j$ 
in order for these operations to represent the fusion rule \cite{Dijkgraaf:1988tf}.
We will come back to this point shortly.

\smallskip

As a concrete illustration, let us consider the simplest case of $\CN=4$ SYM theory, corresponding to
Liouville theory on the torus.
The genus 1 conformal blocks are given by the chiral partition sum,
defined by the trace of $q^{L_0}$ over the sub-space $[V_a]$
\be
\PPsi(a) = {\rm Tr}_{\strut [V_a]} q^{L_0} ,
\ee
with $q=e^{2\pi i \tau}$. These conformal blocks span a linear space, on which the monodromy
operators act. The monodromy operators $\Phi_j(A)$
around the A-cycles manifestly act diagonally,
via eigenvalues that generate the $SU(2)$ representation ring,
and thus are naturally given by (specialized) $SU(2)$ characters. One finds
\be
\Phi_j(A) \, \PPsi(a) \, =\,  W_j(a)\, \PPsi(a)
\ee
with $W_j(a)$ as given in eqn (\ref{wilsonc}).
This establishes the identification of $\Phi_j(A)$ with the Wilson line operators $W_j$.
The action on the conformal blocks generated by
the monodromy around $B$-cycles should reflect the fusion algebra of the corresponding degenerate field \cite{Verlinde:1988sn,Dijkgraaf:1988tf}.
Indeed, one finds that
\be
\Phi_{j}(B) \, \PPsi(a) = \sum_{p=-j}^j\, \PPsi(a + pb ) \label{generalthooft}
\ee
These operators $\Phi_j(B)$ also generate an $SU(2)$ representation ring. Since S-duality interchanges the $A$ and $B$-cycle, we  identify $\Phi_j(B)$ with the 't Hooft loop.
Note that the S-duality map amounts to taking a Fourier-transform, or equivalently:
\be
\Phi_j(B) \, \PPsi(a) \, =\,  W_j(a_D)\, \PPsi(a) \, ; \qquad \quad a_D \equiv \frac{i}{4\pi} \frac{\p\ }{\p a}.
\ee
This relation matches with the results of the semi-classical study in section 2. It is important to note, however,
that it is special to the case of $\CN=4$ SYM theory; in general, the S-duality transformations are much more involved,
as we will see shortly.

We should note here that, to obtain the above results, one needs to apply a standard normalization factor
such that the operator $\Phi_j(B)$ associated with the $B$-cycle, when acting on the identity representation,
produces the degenerate character $\PPsi_j$ with unit pre-coefficient $\Phi_{j}(B)\, \PPsi(0) \spc =\, \PPsi_{j}$ with
$\PPsi_{j} =\!\! \sum\limits_{- {j} \leq \,p \,  \leq j} \!\! \PPsi(\textstyle p\spc b)\, .$
The required normalization factor $\CN_j$ depends on $j$, but is otherwise the same for every path $\gamma$.
To compute the factors $\CN_j$, we perform the monodromy operation
\begin{align}
\inctt{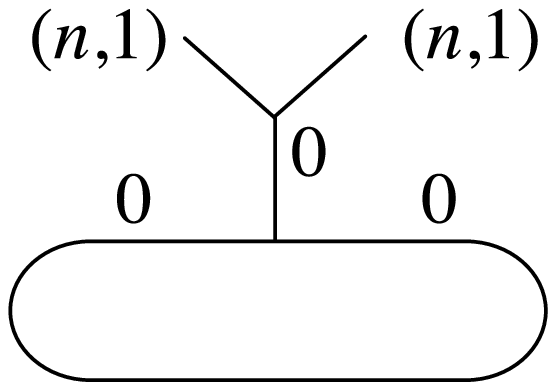} &= F^{-1} \inctt{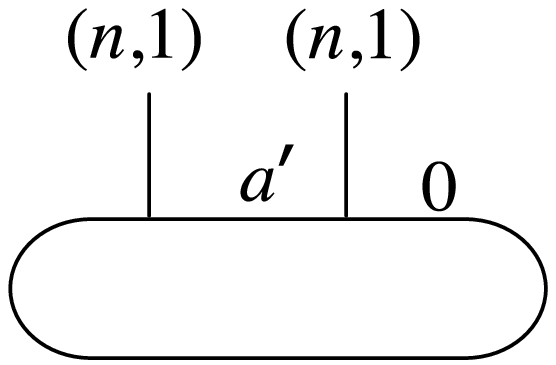}
=  F^{-1}\inctt{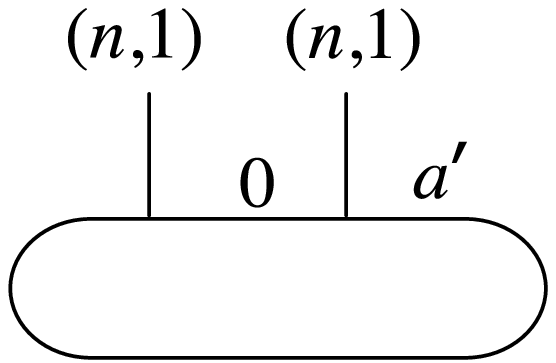} = F^{-1} F\inctt{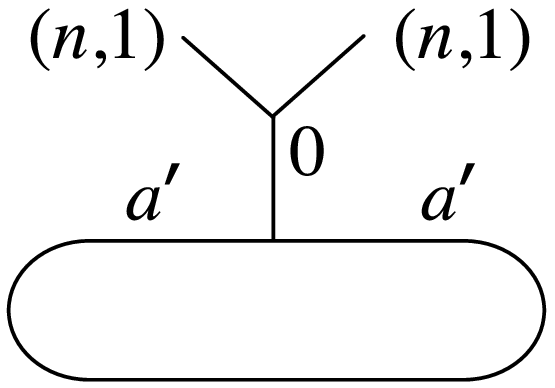} \nonumber
\end{align}
For the first two degenerate insertions under consideration we obtain $\CN_{1/2}= - {\cos \pi b^2},$ and
$\CN_1 = 1+2 \cos(2 \pi b^2).$
In general we expect to find $\CN_j = (-)^{2j} \sum_{-j \leq p\leq j} e^{i \pi p b^2}.$
Upon multiplying the `bare' CFT monodromy operators by this factor, we get the properly
normalized  Verlinde operators $\Phi_j(\gamma)$, that are identified with the 't~Hooft-Wilson loop operators.

In the following we will consider examples of Wilson and 't Hooft loops in the simplest ${\cal N}=2$ gauge theories, namely ${\cal N}=2^*$ and $SU(2)$ with four flavors.

\subsubsection{Example 1: Wilson loop in $N_f=4$}

As a first concrete check, we now compute the Wilson line in the $SU(2)$ gauge theory with $N_f=4$ fundamental flavors, corresponding to the four punctured sphere.

For simplicity, we first focus on the spin 1/2 representation, defined by the monodromy of the degenerate field
$\Phi_{2,1}$. It generates the fusion algebra
\be
[\, \Phi_{2,1}] \, \times \, [ \, V_a \, ]\, = \, [ \, V_{a- \frac{b}{2}} ] + [ \, V_{a+\frac{b}{2}} ]\, .
\ee
According to our previous discussion, we define the {Wilson loop} by the operation:
\begin{align}
\inct{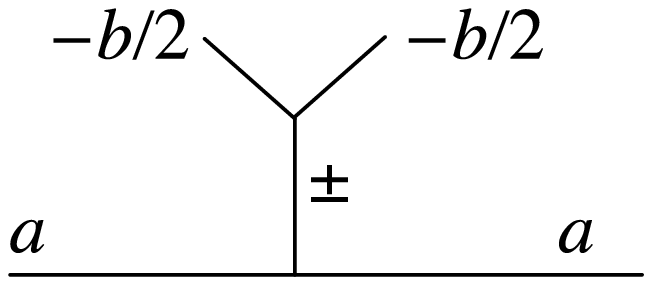}&=F^{-1}\inct{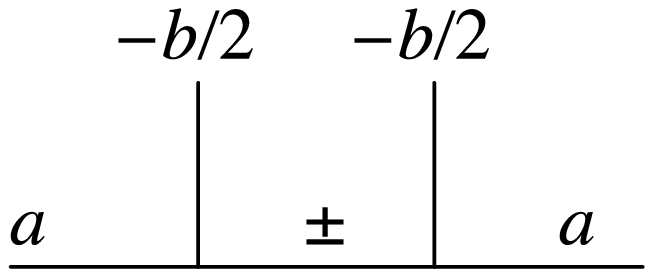}\\
&=F^{-1}\Omega^2\inct{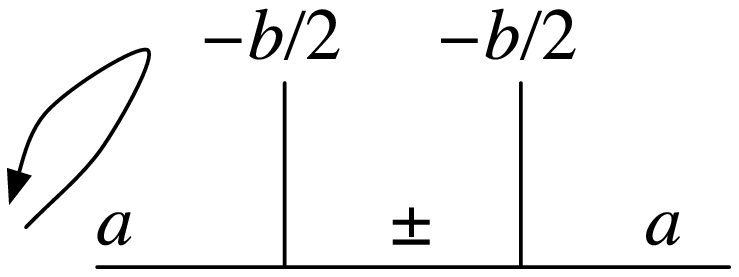}\\
&=F^{-1}\Omega^2 F \inct{w1}
\end{align}
For details on the notation, see Appendix \ref{app:fusionrules}. The degenerate fusion rules imply that the momentum of the intermediate channel of the first and last graph is $-\frac{b}{2} \pm \frac{b}{2}$, which we shortly denote by $\pm$. Notice that the $+$ channel corresponds to the state with zero momentum.
The Wilson loop is then
\begin{equation}
W_{1/ 2} = (F^{-1}\Omega^2 F)_{++}.
\end{equation}
The correct expressions for the fusion and flip matrices can be found in Appendix \ref{app:fusionrules}. Performing the explicit computation
gives
\begin{equation}
W_{1/2}= 2 \cosh(2\pi  b P)\, ;  \label{Wilson} \qquad \quad
a=Q/2 + iP.
\ee
As already mentioned, once the fusion matrices with a degenerate insertion $(2,1)$ are given, we can use the pentagon and hexagon identities in order determine the fusion matrices with a degenerate insertion $(n,1)$.  Via this route,
we have computed the fusion matrices with a degenerate $(3,1)$ insertion.
 We obtain for the corresponding Wilson loop $W_{1}= {1+2 \cosh(4 \pi b P)}\, .$
The general answer can now be guessed, and agrees with the gauge theory result~(\ref{wilsonc})
\begin{equation}
\label{cftwilson}
W_{j}=\sum_{p=-j}^{j}e^{4 \pi p b P}.
\end{equation}

Note that, since the monodromy calculation can be performed locally on a given internal leg of the conformal block, 
this result for the $\CN=2$ gauge theory with $N_f=4$ flavors
is sufficient to fix the form of any Wilson line in any member in our class of $\CN=2$ gauge theories.
The precise match between  (\ref{cftwilson}) and (\ref{wilsonc})
formed the original motivation for our proposed identification of
the Wilson-'t Hooft loop operators
with the Verlinde operators. Combined with the geometric motivation presented in the Introduction,
based on the relation with surface operators, this precise match can be viewed as direct evidence supporting
the conjectured identification between surface operators and degenerate operator insertions in the Liouville CFT.

\subsubsection{Example 2: 't Hooft loop in ${\cal N}=2^*$}

We now turn to the  {'t Hooft loop} of the $\cN=2^*$ theory, corresponding to the torus with one puncture.
It is specified by the following monodromy operation:
\begin{align}
\inct{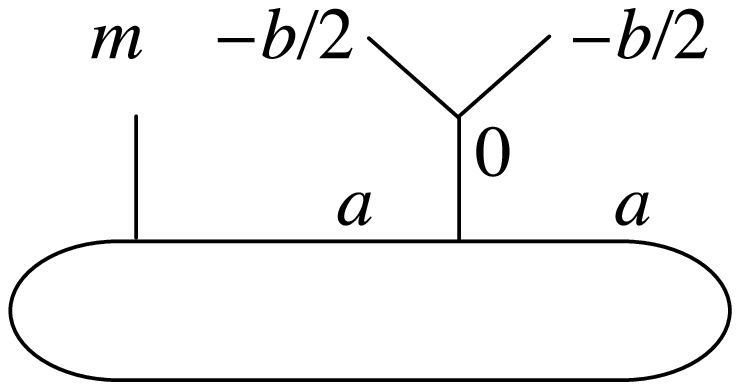} &= F^{-1} \inct{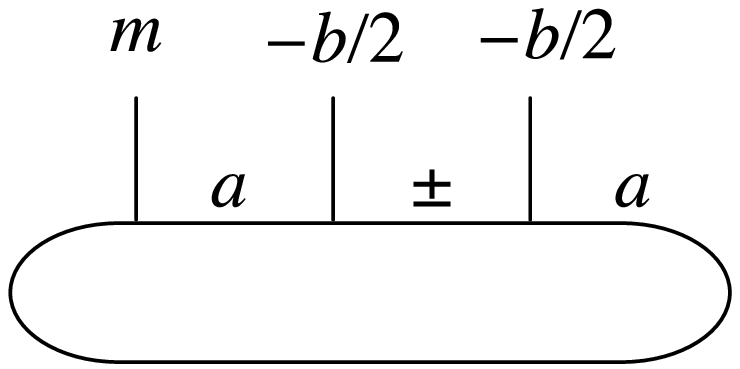} \\
&=  F^{-1} B\inct{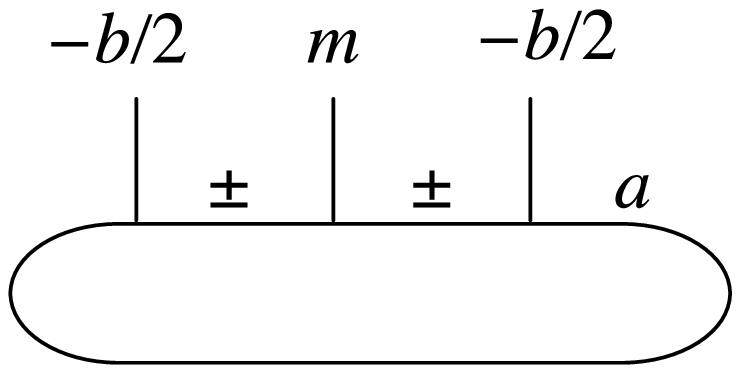} \\
&= F^{-1}  B F\inct{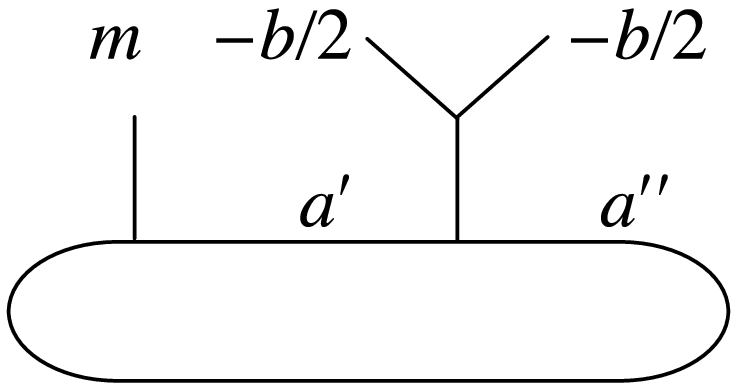}.
\end{align}
Note that in the last line, there are two types of terms:
one has $a'=a''$  and has the vacuum in the fusion of the two degenerate states with $-b/2$;
the other has $a'-a''=\pm b$ and has $-b$ in the fusion of the two states with $-b/2$.
As we did for the case of the Wilson loop, we project on the term which has the vacuum in the fusion. We can write the result in the following form
\begin{equation}
\label{hooftl}
H_{1/2} \, \PPsi(a) =  H_+(a) \PPsi(a+\textstyle \frac 1 2 b) + H_-(a) \PPsi(a-\textstyle \frac 1 2 b)
\end{equation}
Note that the full operation again involves shift operators of the form $e^{\pm \frac{1}{2} {b} \partial_a}$.

In terms of the fusion and braiding matrices
\begin{align}
{H_+(a)} &={\CN_{1/2} }
\left(F\mat{-b/2}{-b/2}{a}{a}\right)^{-1}_{++} B\mat{m}{-b/2}{a}{a+b/2}_{-+}
F\mat{-b/2}{-b/2}{a+b/2}{a+b/2}_{-+} \\[4mm]
{H_-(a)} &= {\CN_{1/2} }
\left(F\mat{-b/2}{-b/2}{a}{a}\right)^{-1}_{+-} B\mat{m}{-b/2}{a}{a-b/2}_{+-}
F\mat{-b/2}{-b/2}{a-b/2}{a-b/2}_{++}
\end{align}
Finally, using the explicit expression for $F$ and $B$ we obtain
\begin{align}
H_+(a)&= 
\frac{\Gamma(2 i b P)\, \Gamma(1+b^2+2 i b P)}
{\Gamma(2i b P+mb)\Gamma(1+b^2+2ibP - mb)},\label{thooft1}\\[3mm]
H_-(a)&= 
\frac{\Gamma(-2 i b P)\, \Gamma(1+b^2-2 i b P)}
{\Gamma(-2i b P+mb)\Gamma(1+b^2-2ibP - mb)}
\label{thooft2}
\end{align}
Here the mass parameter $m$ is normalized so that at 
$m=0$ the theory reduces to $\CN=4$ SYM theory,  i.e. $m=0$ corresponds to inserting the identity operator
at the puncture.
%Note that in $m \to 0$ limit, when ${\cal N}=2^*$ reduces to ${\cal N}=4$,
%we get $H_\pm(a)= 1$.
\bigskip

\noindent
{\it A consistency check}

The Wilson loop operator is ``hermitian,'' in the sense that
inside a full correlation function on $S^4$ one can act with the Wilson 
line either on the holomorphic or the anti-holomorphic
conformal block, and obtain the same result: indeed the Wilson
loop operator (\ref{wilsont}) is diagonal in the integration variable $P$, and
symmetric under $i P \to - i P$.
For consistency, the 't Hooft
loop should satisfy the same constraint.

Up to the usual normalization factor, the integral expression for the $S^4$ expectation value of the
't Hooft loop is (here $b=1$)
\begin{equation}
\label{thooftintegral}
\bigl\langle \, H_{1/2}\, \bigr\rangle_{S^4} = \, \int \! da \, C(a,m,Q-a) \spc \Bigl[\,
\overline{\PPsi(a)}\spc  H_+ (a) \PPsi(a+\textstyle \frac 12 b) \; + \; \mbox{\small $(+\leftrightarrow -)$}\, \Bigr] %c.c.,
\end{equation}
where $C(a,m,Q-a)$ denotes the DOZZ three point function. The DOZZ pre-factor arises as a one loop determinant in the gauge theory, and does not depend on the gauge coupling. The conformal block represent the sum over the classical instanton contributions, and do depend on the gauge coupling. For the $\cN=4$ case,
the partition function on $S^4$ 
further simplifes and reproduces the semi-classical calculation on the gauge theory side performed in \cite{Gomis:2009ir}; for details, see Appendix~\ref{app:nisfourcheck}.

For $\CN=2^*$
the DOZZ three point function takes the form
\begin{equation}\label{N2ZZform}\begin{aligned}
C(a,  m,Q-a)=\left[\pi\mu\gamma(b^2)b^{2-2b^2}
\right]^{-m/b} \frac{\Upsilon_0\Upsilon(2a)\Upsilon(2m)
\Upsilon(2a-Q)}{
\Upsilon(m)^2
\Upsilon(2a+m-Q)\Upsilon(2a-m)}.
\end{aligned}\end{equation}

 The action of the
't Hooft loop is non-diagonal in the integration momentum $P$.
Hence to compare the action of the 't Hooft loop on the
holomorphic conformal block and the action on the
anti-holomorphic conformal block one needs to shift the
integration contour, taking into account the effect of the
shift on the relevant DOZZ three point functions.
A simple calculation shows that the effect of a shift in the integration variable $a$ is:
\begin{equation}\label{ratiop}\begin{aligned}
\frac{C(a+b/2,  m,Q-a-b/2)}{C(a, m,Q-a)}&=
\frac{\gamma(1+b^2+2ib P)
\gamma(2ib P)}{\gamma(2ib P+ m b)
\gamma(1+b^2+2ib P- m b)}.
\end{aligned}\end{equation}
By using the explicit expressions (\ref{thooft1})-(\ref{thooft2}), we recognize the required relation with the 't Hooft loop
coefficients:
\begin{equation}\label{ratiop2}\begin{aligned}
\frac{C(a+b/2,  m,Q-a-b/2)}{C(a, m,Q-a)}= \frac{H_+(iP)}{H_-(iP+b/2)}
\end{aligned}\end{equation}
This relation is sufficient to show that the integral expression when the 't Hooft operator acts on the anti-holomorphic conformal block
coincides with (\ref{thooftintegral}) and hence that  the 't Hooft operator is hermitian.

The DOZZ prefactor can be thought of as part of the integration measure of the integral over the Coulomb branch
parameter $a$. Alternatively, we can choose to absorb it in the definition of the conformal blocks. 
This leads to a somewhat simplified form of the 't Hooft loop expectation values, that suggests that it should be possible to
reproduce the result via a direct gauge theory calculation. We leave this problem for future study.

\subsubsection{Example 3: 't Hooft loop in $N_f=4$}

We can repeat the exercise for the case of $SU(2)$ gauge theory with four flavors. We define the 't Hooft loop in this case by the following operation:
\begin{align}
\inc{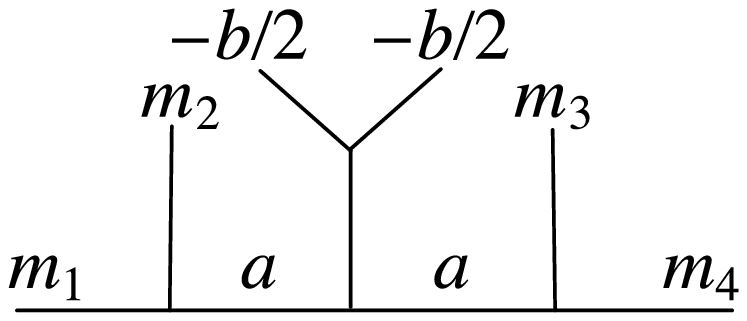}
&=F^{-1} \inc{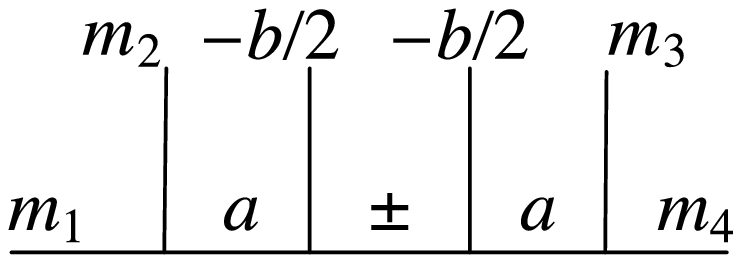}  \label{xx1}\\[1mm]
& = F^{-1} BB\inc{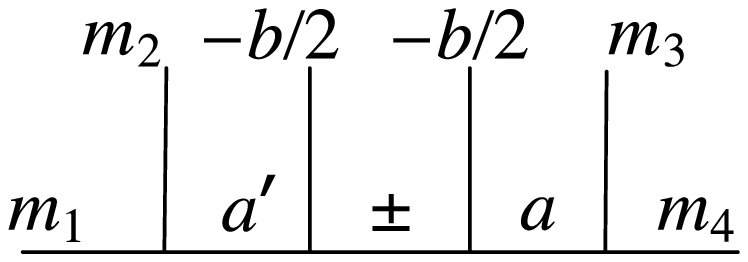} \label{xx2}\\[1mm]
& =F^{-1}  BBB\inc{n3}\label{xx3}\\[1mm]
& =F^{-1} BBBBB\inc{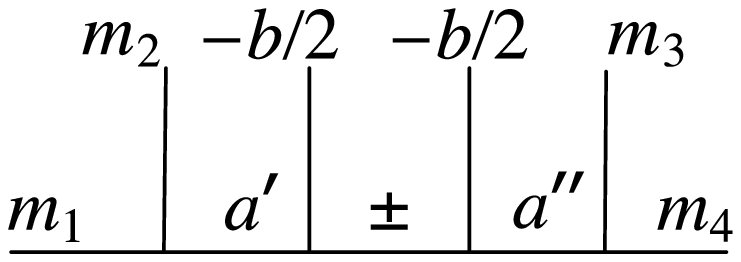} \label{xx4}\\[1mm]
& =F^{-1}  BBBBB F\inc{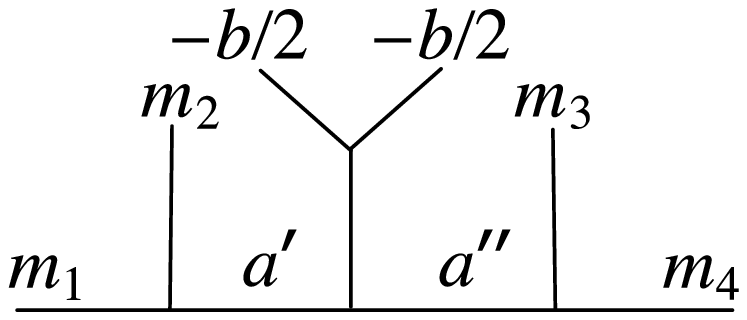}
\end{align}
In words, one braids $m_2$ and $-b/2$ twice, when going from \eqref{xx1} to \eqref{xx2},
then braids $-b/2$ and $-b/2$ when going from \eqref{xx2} to \eqref{xx3},
then braids $-b/2$  and $m_3$ twice when going from \eqref{xx3} to \eqref{xx4}.
In the last line, $a'$ and $a''$ can take the values $a$ and $a \pm b$. As before, we project on the channel $a'=a''$.
Using the explicit expressions for the fusion and braiding
matrices in the appendix, we obtain the following
result\footnote{We removed an overall phase factor
$e^{\frac{3}{2} i \pi b^2}$. This type of phase factor can be
produced, say, by an extra braiding move of one degenerate
field around the other in the vacuum channel. Such spurious $P$
independent phase factors are subtle to track down across
S-duality, unless one goes carefully through the full set of
algebraic manipulations in the Moore-Seiberg groupoid. More
simply, we remove it here by requiring $H_0$ to be real.}
\begin{equation} \label{hooftl2} H_{1/2} \, \PPsi(a) =  H_+(a)
\PPsi(a+ b) + H_0(a) \PPsi(a) +  H_-(a) \PPsi(a- b)
\end{equation}
with
\begin{eqnarray}
\label{hpm}
H_{\pm}(a) =-\frac{2 \pi^2 {\rm csc}( \pi b^2)}{\Gamma[-b^2]\Gamma[1+b^2]} \times\hspace{9cm} \\[1mm]
{\Gamma[1+2b(b\pm i P)]\Gamma[b(b\pm 2i P)]\Gamma[\pm 2i b P]\Gamma[1+b^2\pm 2 i b P] \over \prod\limits_{s_i=\pm} \Gamma[{1 \over 2}(1+b^2 \pm 2 i b(P+s_1 m_1+s_2 m_2))]\Gamma[{1 \over 2}(1+b^2\pm 2 i b(P+s_3 m_3+s_4 m_4))]} \nonumber
\end{eqnarray}
and\\[-8mm]
\begin{eqnarray} %multline}
\label{hnot}
H_0(a)  =
\frac{4 \cos \pi b^2 }{ \cosh 4\pi b P - \cos 2\pi b^2}%[\sum e^{2b(\pm m_2\pm m_3)\pi}+e^{2b(\pm m_1\pm m_4)\pi}]
(\cosh 2\pi b m_2 \cosh 2\pi bm_3 + \cosh 2\pi b m_1 \cosh 2\pi bm_4)\nonumber \\[-1mm] \\[-1mm]
+\frac{4 \cosh 2\pi b P}{ \cosh 4\pi b P - \cos 2\pi b^2}
(\cosh 2\pi b m_1 \cosh 2\pi bm_3 + \cosh 2\pi b m_2 \cosh 2\pi bm_4)\, .\nonumber
%[\sum e^{2b(\pm m_1\pm m_3)\pi}+e^{2b(\pm m_2\pm m_4)\pi}]}{ \cosh 4\pi b P - \cos 2\pi b^2}
\end{eqnarray} %{multline}
The above formulas are clearly more complicated than those of the $\cN=2^*$ theory, and it would seem to be a
true challenge to reproduce them via a direct gauge theory calculation. However, we expect that, similar as for $\cN=2^*$,
the prefactors $H_\pm(a)$ can be considerably simplified absorbing the DOZZ factor/one-loop determinant
into the definition of the conformal blocks. The diagonal factor $H_0(a)$, however, can not be simplified in this way.

Note further that, although the above 't Hooft operator
is associated %to the (2,1) degenerate field, and thus
to the spin 1/2 representation of $SU(2)$, its
action on the chiral partition functions looks more like that of a spin 1 loop operator, at least when compared to
the $\CN=2^*$ answer (\ref{hooftl}). The geometric reason for this is that to perform the monodromy operation for
the $N_f=4$ theory, the degenerate insertion needs to pass the intermediate leg of the conformal block twice.
The physical reason is that the 't Hooft operator with minimal magnetic charge in the $N_f=4$ theory, which has fields in the doublet of the gauge group,
has twice the magnetic charge of the minimal 't Hooft loop of the $\cN=2^*$ theory,
which has fields only in the adjoint of the gauge group.\footnote{This fact was already noted in \cite{Kapustin:2006hi}.}

The S-duality relation between Wilson and 't Hooft line operators can be explicitly demonstrated,
by standard manipulations in the Moore-Seiberg groupoid. The main part of the computation
was already done in section 4.2 for the corresponding line operators acting on a surface operator.
The new ingredients are the initial and final fusions from and to the identity, which add little
extra complication.

\def\PP{\mbox{\small \spc P}}
\def\XX{{\rm X}}

As a more conceptual point, we observe that the linear action of the Wilson-'t Hooft loops on the chiral partition
function is independent of the gauge coupling~$\tau$. This is of course an automatic consequence of the fact that both
are specified as elements of the Moore-Seiberg groupoid of the Liouville CFT, which is generated by fusion and
braiding matrices that do not depend on the complex structure of Riemann surface $C$. This motivates us to look
for a  more intrinsic formulation of the loop operators, in which this independence is more manifest.

\medskip

\subsection{Loop operators from quantum Teichm\"uller space}

The modular geometry of Liouville CFT  identifies the space of conformal blocks with a linear representation
space on which the fusion and braiding matrices and the loop operators act as a non-commutative set of
unitary and hermitian operators, respectively. It is thus natural to expect that there should exist a suitable phase
space that after quantization yields the Liouville conformal blocks as Hilbert states.
The Verlinde operators would then be given by suitable functions, defined on this phase space.
For the Liouville-Virasoro conformal blocks associated with some genus $g$ Riemann surface $C$ with $n$ punctures,
there is a natural candidate for such a phase space:  the Teichm\"uller space $\CT_{g,n}$ of $C$. This relation between
Liouville CFT and  the quantization of $\CT_{g,n}$ was conjectured in \cite{Verlinde:1989ua}, and recently proven in \cite{Teschner:2005bz}.

Teichm\"uller space $\CT_{g,n}$ can be thought of as the space of constant curvature
metrics on $C$. This also happens to be space of classical solutions to the Liouville equations on $C$.
$\CT_{g,n}$ is known to be $6g-6+ 2n$ dimensional symplectic manifold, with
symplectic form given by the Weil-Peterson form. It can thus be quantized.

A convenient set of observables is obtained as follows. We
may specify the  constant curvature 2-d metric via a zweibein and a spin connection, which in turn combine into a flat $SL(2,\IR)$ gauge field $A$.
To any (non-self-intersecting) path $\gamma$ on $C$, we can thus associate the Wilson-like loop $L(\gamma) = {\rm tr}_{\frac 12} \exp \oint_\gamma A$, where the trace is taken in the spin 1/2
representation of $SL(2,\IR)$. $L(\gamma)$ can be expressed in terms of
the geodesic length $\ell(\gamma)$ of~$\gamma$ via
\be
\label{Lloop}
L(\gamma) = 2\cosh 2\ell(\gamma)\, .
\ee
In the quantized theory, these operators in general only commute
in case the corresponding curves do not intersect. We can thus define a maximally commuting set of observables, by choosing the set of $L(\gamma)$'s for all the dividing cycles of a pant decomposition of the Riemann surface $C$.
The Hilbert space of the quantum theory is thus naturally labeled by the eigen values of this maximally
commuting set of operators.

This structure is of course reminiscent of the way the Liouville CFT loop operators $\Phi_j(\gamma)$ act on the
space of conformal blocks.
In fact, we claim that the operators $L(\gamma)$ can be identified with the Verlinde
monodromy operators of
the lowest degenerate field $\Phi_{2,1}$.
\be
\Phi_{\frac 12}(\gamma) \equiv  L(\gamma)\, .
\ee
A semi-classical motivation for this identification is that the degenerate field equation
(\ref{degendiff})  tells us that moving $\Phi_{2,1}$ proceeds via parallel transport via a flat $SL(2,\IR)$ connection $A = \left(\begin{array}{cc} 0 \, & b^2 T \\ 1 \,& 0\end{array} \right)$ in the spin 1/2 representation. In the full quantum theory,
the result was established in \cite{Teschner:2005bz}. The same type of argument can be generalized to the
degenerate operators $\Phi_{2j+1,1}$, to show that the corresponding monodromy operation can be identified with
the spin $j$ Wilson loop  $\tr_j \exp \oint_\gamma A =  \!\! \sum\limits_{-j\leq p\leq j}\!\!  e^{2p \ell(\gamma)}$.

As concrete illustration, we return to the $N_f=4$ example discussed in the subsection 4.2.3. This case was analyzed in detail in \cite{Teschner:2002vx}. We will not repeat his analysis here, but only state the main results relevant to our problem
of computing the action of the loop operators. The most convenient construction of the quantized Teichm\"uller
theory proceeds via the introduction of so-called Fock variables. In the case of the four punctured sphere,
these comprise a single pair  of canonically conjugate variables $\hat\PP$ and $\hat{\mbox{\small X}}$, with
$[\spc \hat\PP, \hat{\mbox{\small X}}\spc ] = i.$
The A and B-cycle operators are expressed in terms of   $\hat\PP$ and $\hat{\mbox{\small X}}$ as  \cite{Teschner:2002vx} (here for simplicity,  we assume that all mass parameters $m_i$ are equal)
\begin{align}
\hat{L}(A) & = 2 \cosh(2 \pi b \hat \PP) + e^{- \frac 12  b   \hat \XX} \bigl[ 4 \cosh^2 (\pi b \hat \PP) \bigr] e^{- \frac 12 b \hat \XX}
\nonumber \\[-1mm]
\label{LAB}
\\[-1mm]
\hat{L}(B) & = 2 \cosh(2 \pi b \hat \PP) + e^{  \frac 12 b   \hat \XX} \bigl[ 4 \cosh^2(\pi b (\hat \PP-m)) \bigr] e^{\frac12 b  \hat \XX}
\nonumber
\end{align}
These two loop operators do not commute when $b \neq 1$, but for $b=1$, they do commute. We will comment on this
distinction in the concluding section.\footnote{For general $b$, there exists a natural dual pair of operators $\tilde{L}(A)$ and
$\tilde{L}(B)$ given by the same expressions (\ref{LAB}), with $b$ replaced by $1/b$. The first pair (\ref{LAB})
represent the monodromy loops of the degenerate field $\Phi_{2,1}$ and the dual pair represent the monodromy loops
of $\Phi_{1,2}$. The two dual pairs of operators commute with each other, but not among each other.}

The  conformal block with fixed Liouville momentum along the intermediate channel is now identified with the
eigen state $|\Psi_a\rangle$ of the A-cycle operator $\hat L(A)$, with eigenvalue
\be
\hat{L}(A)\spc | \Psi_a \rangle = 2\cosh(2\pi b P) \, |\Psi_a\rangle\, .
\ee
These eigen states have been explicitly constructed in \cite{Ponsot:2001ng}.
Via the gauge theory Liouville correspondence,
$|\Psi_a\rangle$ represents the Nekrasov partition sum with Coulomb parameter $a$, and $\hat L(A)$ is the spin $\frac 12$
 Wilson line.
Note however that the quantum system has been defined independent of the gauge coupling constant $\tau$, and hence, without introducing extra structure, it can not be used to compute gauge theory quantities that depend on $\tau$.

The spin $\frac 12$  't Hooft loop is found by computing the action of the dual loop operator $\hat{L}(B)$ on the
eigen states of $\hat{L}(A)$
\be
\hat{L}(B)\spc | \Psi_a \rangle= H_+(a) | \Psi_{a+ b}\rangle + H_0(a) |\Psi_a\rangle +  H_-(a) |\Psi_{a- b}\rangle\, .
\ee
The results of \cite{Teschner:2002vx} imply that, in a suitable normalization of $|\Psi_a\rangle$, the above pre-factors
$H_\pm(a)$ and $H_0(a)$ coincide with the results (\ref{hpm}) and (\ref{hnot}) found from the Liouville CFT.
The fusion matrix ${\cal F}$ that implements S-duality of the $N_f=4$ theory, is the unitary basis transformation that
relates the eigen states of $\hat{L}(A)$ and $\hat{L}(B)$.

To conclude, we learn that the Wilson-'t Hooft loop operators, when acting on the Nekrasov partition functions form a
non-commutative ring, given by the ring of functions (\ref{Lloop}) on the quantized Teichm\"uller space.

\section{Conclusion}

In this paper we have studied some basic properties of surface and loop operators in a class of $\CN=2$ $SU(2)$ quiver gauge theories, obtained by compactifying the 6-d $(2,0)$ theory on a Riemann surface $C$. We have exploited
the identification of the instanton partition sum of the gauge theory with the conformal blocks of Liouville CFT, to
define the expectation values of the surface and loop operators in terms of natural quantities in the CFT.
In the 2-d CFT formalism, non-perturbative properties of the gauge theory, such as S-duality, can be made manifest.
We end with some comments on our main results, and point to some important open problems.

We have found that the Wilson-'t Hooft line operators are naturally represented via a non-commutative ring
of linear operators $\Phi(\gamma)$,
that act on the instanton partition functions of the $\epsilon$ deformed theory on $\IR^4$.
This raises a small basic puzzle, since in general,  there is no natural way to define a
commutation relation between line operators on $\IR^4$.  However, in the $\epsilon$-deformed theory there
is a special supersymmetric sub-class of loop operators that are left invariant under the $U(1)$ symmetry
(\ref{torusact}). For $b = \sqrt{\frac{\epsilon_1}{\epsilon_2}} \neq 1 $ the invariant loops must be
located at $w_1=0$ or at $w_2=0$.
When restricted to each of these 2-d subspaces, loop operators do allow a natural time ordering,
e.g. by using the radial time coordinate $\exp(t) = |w_1|^2 + |w_2|^2$.
The loop operators thus may represent a non-commutative ring.
On the other hand, in the special case that $b=1$,
the $U(1)$ symmetry (\ref{torusact}) leaves invariant a continuous family of circular loops,
given by $t =const.$ within any plane of the form $c_1 w_1 + c_2 w_2 = 0$.
When acting at the same radial time $t$, two such circular loops in different planes are automatically linked.
Locality restricts the commutation relation between linked loop operators to elements of the center of the
gauge group $G$. In the case of the $SU(2)$ quiver theories,  two different loop operators must therefore
either commute or anti-commute for $b=1$. This is indeed the case for our construction.
The two operators in the $N_f=4$ theory given in (\ref{LAB}) are a specific example: it is easy to check that in this case
the 't Hooft loop commutes with the Wilson loop.

\smallskip

Perhaps the most important lesson from our study is that it has illustrated the central role played by
the surface operators of the supersymmetric gauge theory.
It is evident that surface operators have a very rich set of properties,
that are well worth analyzing in much more detail.
%and in this paper we have only scratched their surface.
 In particular, it would be a most useful advance if one
could establish our conjectured identification with a local degenerate field placed at
a point on the Riemann surface $C$. One possible route is to try to make contact with the
work of Braverman \cite{Braverman:2004vv}, who has independently proposed that the instanton
partition function in the presence of a surface operator should
satisfy a differential equation of the type (\ref{degendiff}).

\section*{Acknowledgments}
We have benefited from useful discussions with N. Drukker, J. Gomis, J. Maldacena,  N. Nekrasov, T. Okuda,
V. Pestun, N. Seiberg, J. Teschner, C. Vafa,  E. Verlinde, and E. Witten.
L.F.A. and D.G. are supported in part by the DOE grant DE-FG02-
90ER40542. D.G. is supported in part by the Roger Dashen membership in the Institute for Advanced
Study. YT is supported in part by the NSF grant PHY-0503584, and by the Marvin L.
Goldberger membership at the Institute for Advanced Study. The research of H.V. is supported
by the National Science Foundation under Grant No. PHY-0756966 and by an Einstein
Fellowship of the Institute for Advanced Study. The work of SG is supported in part by DOE grant DE-FG03-92-ER40701, in part by NSF grant PHY07-57647, and in part by the Alfred P. Sloan Foundation. Opinions and conclusions expressed here are those of the authors and do not necessarily reflect the views of funding agencies.

\bigskip

\noindent
{\bf Note added}\qquad 
While in the process of writing up our results, we were informed of related work done by N. Drukker, J. Gomis, T. Okuda and
J. Teschner. We coordinated the time of release of our paper with theirs, \cite{Drukker:2009id}.

\appendix

\section{Useful formulae}

We start by defining the Barnes double Gamma function. Barnes double zeta function is defined as
\begin{equation}
\zeta_2(s;x|\epsilon_1,\epsilon_2)=\sum_{m,n} (m\epsilon_1+n\epsilon_2+x)^{-s}
=\frac{1}{\Gamma(s)} \int_0^\infty \frac{dt}{t} t^s
\frac{e^{-tx}}{(1-e^{-\epsilon_1 t})(1-e^{-\epsilon_2 t})}
\end{equation}
from which Barnes' double-Gamma function is defined as
\begin{equation}
\Gamma_2(x|\epsilon_1,\epsilon_2)=
\exp \frac{d}{ds}\Bigm|_{s=0}\zeta_2(s,x|\epsilon_1,\epsilon_2).
\end{equation} The arguments $\epsilon_{1,2}$ in $\Gamma_2$ will be
often omitted if there is no confusion. Assume $\epsilon_{1,2}\in \mathbb{R}_{>0}$.
Then Barnes' double-Gamma function
is analytic in $x$ except at the poles at $x=-(m\epsilon_1+n\epsilon_2)$
where $(m,n)$ is a pair of non-negative integers.
Therefore one can think of Barnes' double-Gamma as the
regularized infinite product
\begin{equation}
\Gamma_2(x|\epsilon_1,\epsilon_2) \propto \prod_{m,n \ge 0}
\left(x+m\epsilon_1+n \epsilon_2\right)^{-1}.\label{infiniteproduct}
\end{equation}
%Furthermore it is real when $x$ is real.
%As such, \begin{equation}
%\Gamma_2(x^*)=
%\Gamma_2(x)^*.
%\end{equation}
%Another relation we need is \begin{equation}
%\Gamma_2(x+\epsilon_1)\Gamma_2(x+\epsilon_2)
%=x \Gamma_2(x)\Gamma_2(x+\epsilon_1+\epsilon_2).\label{gamma-shift}
%\end{equation}
%This is a natural property the infinite product in the right hand side
%of \eqref{infiniteproduct} would have.
An important property is \begin{equation}
\frac{\Gamma_2(x+\epsilon_1|\epsilon_1,\epsilon_2)}
{\Gamma_2(x|\epsilon_1,\epsilon_2)} =
\frac{\sqrt{2\pi}}{\epsilon_2{}^{x/\epsilon_2-1/2} \Gamma(x/\epsilon_2)} \label{gammagamma}
\end{equation}

The three point function of primaries in Liouville theory is given by the DOZZ formula in terms of
\begin{equation}\label{ZZform}\begin{aligned}
C(\alpha_1, & \alpha_2,\alpha_3)=\left[\pi\mu\gamma(b^2)b^{2-2b^2}
\right]^{(Q-\sum_{i=1}^3\alpha_i)/b}\times\\
& \times \frac{\Upsilon_0\Upsilon(2\alpha_1)\Upsilon(2\alpha_2)
\Upsilon(2\alpha_3)}{
\Upsilon(\alpha_1+\alpha_2+\alpha_3-Q)
\Upsilon(\alpha_1+\alpha_2-\alpha_3)\Upsilon(\alpha_2+\alpha_3-\alpha_1)
\Upsilon(\alpha_3+\alpha_1-\alpha_2)}.
\end{aligned}\end{equation}
where the $\Upsilon$ and $\gamma$ functions are given by
\begin{equation}\label{updef}\begin{aligned}
{} & \up_b(x)=
\frac{1}{\Ga_b(x)\Ga_b(Q-x)},\qquad \Ga_b(x)=\Ga_2(x|b,b^{-1}),\qquad \gamma(x)=\Gamma(x)/\Gamma(1-x) \end{aligned}\end{equation}

\section{Fusion and braiding} \label{app:fusionrules}
In this appendix we briefly review the definition of fusion and braiding matrices and the identities they satisfy. We will follow closely the review \cite{Moore:1989vd} to which we refer the reader for more details. For Liouville theory, the relevant
results were developed in  \cite{Teschner:2003en} and references therein. 
We use the following pictorial representation for the conformal block of the four-point function
\begin{equation}
%{\cal F}_a^{a_1,a_2,a_3,a_4} \sim
\Bigl \langle  V_{a_1}(0) V_{a_1}(1) V_{a_3}(\infty) V_{a_4}(q)\Bigr\rangle_{\{ a\} } \; = \;
\inc{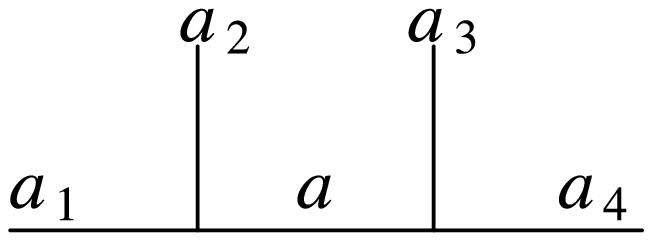},
\end{equation}
where $a_i$ denote the Liouville momenta of the states. Usually these momenta are chosen to lie in the physical line $a_i={Q \over 2}+i P_i$, with $P_i$ real, however, sometimes, as discussed in detail bellow, we will consider "degenerate" values of the form $i P={-n \over 2}b-{m \over 2 b}$. Fusion and braiding matrices are defined as the ones linearly relating different sets of blocks, for instance
\begin{equation}
\inc{f1}=\sum_{a'} F_{aa'}\mat{a_2}{a_3}{a_1}{a_4} \inc{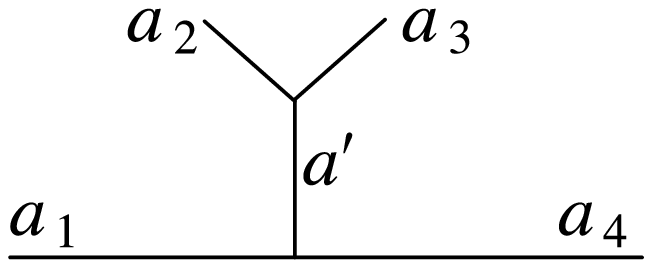}
\end{equation}
In principle, the index $a'$ could run over a continuous set, however for the case considered in this paper we will focus on discrete sums, as argued bellow. We will often choose one the external states, lets say $a_2$, to be the degenerate field $V_{2,1}$, namely $a_2=Q/2+{-2 \over 2}b-{1 \over 2 b}=-b/2$. In this case, the "degenerate" fusion rules imply that $a=a_1 \pm b/2$ and $a'=a_3 \pm b/2$. Another case of interest is the case of the identity operator, in which $a_2=0$, in this case $a=a_1$ and $a'=a_3$. The fusion matrix has the following symmetries
\begin{equation}
F_{aa'}\mat{a_2}{a_3}{a_1}{a_4}=
F_{aa'}\mat{a_1}{a_4}{a_2}{a_3}=
F_{aa'}\mat{a_3}{a_2}{a_4}{a_1}.
\end{equation}
In addition, the fusion matrices satisfy the following orthogonality conditions
\begin{equation}
\sum_{a'} F_{aa'}\mat{a_2}{a_3}{a_1}{a_4}
F_{a'a''}\mat{a_2}{a_1}{a_3}{a_4}
=\delta_{a a''}.
\end{equation}
In a similar manner, we define the braiding matrices
\begin{equation}
\inc{f1}=\sum_{a'} B_{aa'}^{(\epsilon)}\mat{a_2}{a_3}{a_1}{a_4} \inc{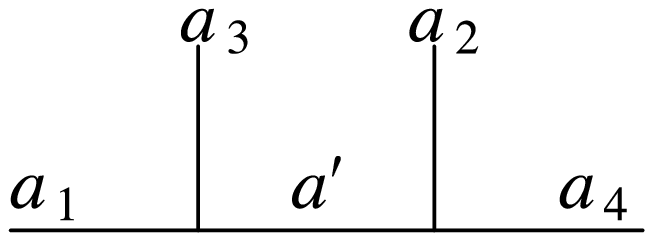}
\end{equation}
where $\epsilon =\pm 1$ denotes the sense of the braiding. $B$ and $F$ satisfy the following relation \begin{equation}
F_{aa'}\mat{a_2}{a_3}{a_1}{a_4}
=e^{-\epsilon i\pi (\Delta_{a_1}+\Delta_{a_3}-\Delta_{a}-\Delta_{a'})}B_{aa'}^{(\epsilon)}\mat{a_2}{a_4}{a_1}{a_3}
\end{equation}
A particular case of the braiding matrix, in which of the the external states is the identity, flips $a_1$ and $a_2$, defining the flip operator \begin{equation}
\inc{f1}= \Omega(\epsilon)^a_{a_1,a_2}\inc{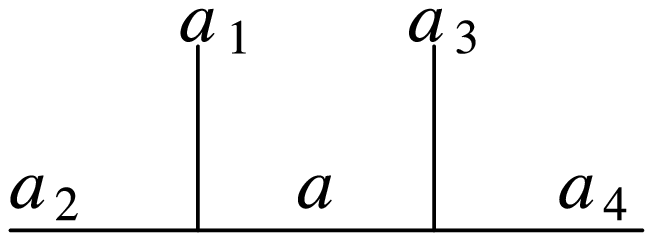}
\end{equation} where \footnote{Note that if one of the entries of $\Omega$ is degenerate, then the other two entries should be related, so in this case the corresponding $\delta$-function is missing from our definition of $\Omega$.}
\begin{equation}
\Omega(\epsilon)^a_{a_1,a_2}=e^{\epsilon\pi i (\Delta_{a_1}+\Delta_{a_2} - \Delta_a) }
\end{equation}
and $\Delta_a=a(Q-a)$ is the conformal dimension of the given operator.

Fusion and braiding matrices are known to satisfy several identities \cite{Moore:1989vd} \cite{Teschner:2003en}. 
In particular they satisfy the so-called pentagon and Yang-Baxter identities
\begin{eqnarray}
\sum_s F_{p_2 s}\mat{j}{k}{p_1}{c}F_{p_1 l}\mat{i}{s}{a}{c}F_{sr}\mat{i}{j}{l}{k}&=&F_{p_1 r}\mat{i}{j}{a}{p_2}F_{p_2l}\mat{r}{k}{a}{c}\\[3mm]
\sum_p B_{a_6 p}^{(\epsilon)}\mat{a_2}{a_3}{a_1}{a_7} B_{a_7 a_9}^{(\epsilon)}\mat{a_2}{a_4}{p}{a_5}B_{p a_8}^{(\epsilon)}\mat{a_3}{a_4}{a_1}{a_9}&=&\sum_p B_{a_7 p}^{(\epsilon)}\mat{a_3}{a_4}{a_6}{a_5} B_{a_6 a_8}^{(\epsilon)}\mat{a_2}{a_4}{a_1}{p}B_{p a_9}^{(\epsilon)}\mat{a_2}{a_3}{a_8}{a_5} \nonumber
\end{eqnarray}
The so-called hexagon identity is obtained from the Yang-Baxter relation by setting, lets say, $a_5=0$ and using the fact that $F_{a_1 a_3}\mat{0}{a_3}{a_1}{a_4}=1$.

\bigskip

\subsection{Degenerate fusion} \label{app:degfusion}
The Liouville theory correlation functions are naturally defined for normalizable
vertex operators, whose Liouville momentum $\alpha$  lies on the physical line $\alpha=\frac{Q}{2} + i P$ for real $P$, i.e. with conformal dimensions greater than $\frac{Q^2}{4}$.
It is sometimes useful, though, to analytically continue such correlation functions to other values of the momenta,
especially to the degenerate values $i P=-\frac{n}{2} b - \frac{m}{2b}$. A correlation function with one degenerate field satisfies a holomorphic differential equation due to the presence of a null vector in the Verma module of the degenerate field.

A correlation function with all momenta on the physical line can be decomposed into conformal blocks and written as a
multiple integral over the momenta on the internal legs, which also lie on the physical line.
As the conformal blocks are analytic in the conformal dimensions, and satisfy individually the
Ward identities for the energy momentum tensor, one may imagine that the conformal blocks with a degenerate insertion will also satisfy the same differential equation as the full correlation function. However, this is not the case,
essentially because the insertion of a null vector does not make the conformal block automatically zero. Rather,
the conformal block vanishes identically if and only if  the internal/external  momenta adjacent to the
insertion of the null field are analytically continued to values satisfying the degenerate fusion relations
$i P_1 - i P_2 = r b - \frac{s}{b}$ where $r,s$ are the weights of $SU(2)$ representations of spin $\frac{n-1}{2}$ and $\frac{m-1}{2}$ respectively. Hence only conformal blocks satisfying this constraint will satisfy the differential equation.

Correspondingly, if one of the external Liouville momenta in a correlation function is analytically continued to
a degenerate value, we expect one of the integrals over continuous, physical momenta to ``localize'' to a discrete sum
over the values allowed by the degenerate fusion relations. The mechanism is rather simple and can be understood from the form of the DOZZ three point function.
As a function of the external momenta, say $\alpha_1$, the DOZZ three point function has actually a zero at all degenerate values, because of the factor $\Upsilon(2\alpha_1)$ in the numerator. In the full correlation function, however, this zero is compensated by a crucial divergence produced by the analytic continuation: the poles produced by the factors $\Upsilon(\alpha_1+\alpha_2-\alpha_3)\Upsilon(\alpha_3+\alpha_1-\alpha_2)$ at $\alpha_3 = \alpha_2 + (\alpha_1+ n b + m b^{-1})$ and at  $\alpha_3 = \alpha_2 - (\alpha_1+ n' b + m' b^{-1})$ move towards each other and end up pinching the integration path as $\alpha_1$ passes through the degenerate values. One can deform away the path from the pinching poles, say keeping it in the canonical region for one $\Upsilon$ function, at the price of collecting extra residues as the poles of the other $\Upsilon$ function are crossed. These poles satisfy the degenerate fusion relations for $\alpha_1, \alpha_2$. An almost identical contribution comes from the other two $\Upsilon$ functions at the denominator.

As a result, we are left with a sum over the residues, i.e. conformal blocks which satisfy the degenerate fusion constraints, and hence the differential equation. We will call these conformal block ``degenerate conformal blocks''

Any sort of fusion and braiding operations in the presence of a degenerate field must send
solutions of the differential equation to solutions of the differential equation.  Hence the action of a fusion operation on a degenerate conformal block should give a combination of degenerate conformal blocks, rather than an integral over all possible momenta in the intermediate channels. The mechanism is presumably similar as the one for the correlation functions. There are two integrals: one in the definition of the fusion matrix itself, and one over the internal momentum of the fused conformal block. The fusion matrix has a zero both when an external momentum becomes degenerate, {\em and} then furthermore when the internal momentum satisfies the degenerate fusion rule. The zeros will kill the continuum contributions, and spare only the discrete residues accumulated during the analytic continuation of the external momentum to the degenerate value, and of the
internal momentum to the value dictated by the degenerate fusion relation.

As we will see below, we do not need to compute those residues: the fusion matrix involving
a degenerate field used in this paper
can be extracted directly from the explicit solutions to the differential equations for four point
degenerate conformal blocks on the sphere.

For the computations relevant to this paper, we only need the answer for the simplest case, the $(2,1)$ degenerate field.
The genus zero four point correlation functions with a $(2,1)$ insertion satisfy a degree $2$ differential equation which reduces to the hypergeometric equation. The equation and solutions are actually determined uniquely by the behavior as the degenerate puncture, located at the point $q$, approaches
the other punctures at $0,1,\infty$ of Liouville momenta $\alpha_{0,1,\infty}$. The two conformal blocks in the $s$ channel, with internal momentum $\alpha_0 \pm \frac{b}{2}$ are
\begin{eqnarray}
{\PPsi}_-^s & = & q^{\alpha_0 b} (1 - q)^{\alpha_1 b} \times \nonumber \\
  & & {}_2F_1\left( (\alpha_0 + \alpha_1 + \alpha_{\infty}-\frac{b}{2}-Q) b, (\alpha_0 + \alpha_1 - \alpha_{\infty}-\frac{b}{2}) b, (2 \alpha_0 - b) b; q\right)\nonumber\\[2mm]
{\PPsi}_+^s & = & q^{(Q - \alpha_0) b} (1 - q)^{\alpha_1 b} \times \nonumber \\
  & & {}_2F_1\left((Q - \alpha_0 + \alpha_1 - \alpha_{\infty} - \frac{b}{2})b, (-\alpha_0 + \alpha_1 +
   \alpha_{\infty} - \frac{b}{2})b, (2Q-2 \alpha_0 - b) b; q\right)\nonumber
\end{eqnarray}

The s-channel conformal blocks can be rewritten in terms of t-channel conformal blocks
by standard hypergeometric identities, from which the fusion coefficients can be computed
\begin{align}
F_{--}\mat{-b/2}{a_3}{a_1}{a_4}&=
\frac{\Gamma[(2a_1-b)b]\,\Gamma[(Q-2a_3)b]}
{\Gamma[(a_1-a_3+a_4-\frac{b}{2})b]\,\Gamma[1+(a_1-a_3-a_4)b+b^2/2]}\\[2mm]
F_{-+}\mat{-b/2}{a_3}{a_1}{a_4}&=
\frac{\Gamma[(2a_1-b)b]\,\Gamma[(2a_3-Q)b]}
{\Gamma[(a_1+a_3-a_4-\frac{b}{2})b]\,\Gamma[(a_1+a_3+a_4-\frac{b}{2}-Q)b]}\\[2mm]
F_{+-}\mat{-b/2}{a_3}{a_1}{a_4}&=
\frac{\Gamma[1+(Q-2a_1)b]\,\Gamma[(Q-2a_3)b]}
{\Gamma[1-(a_1+a_3-a_4-\frac{b}{2})b]\,\Gamma[1-(a_1+a_3+a_4-\frac{b}{2}-Q)b]}\\[2mm]
F_{++}\mat{-b/2}{a_3}{a_1}{a_4}&=
\frac{\Gamma[1+(Q-2a_1)b]\,\Gamma[(2a_3-Q)b]}
{\Gamma[(-a_1+a_3+a_4-\frac{b}{2})b]\,\Gamma[1-(a_1-a_3+a_4)b+b^2/2]}
\end{align}
It is straightforward to verify the basic pentagon/Yang-Baxter/hexagon identities which involve such degenerate fusion matrices only.\footnote{These relations can also be used to produce the
fusion matrices for higher degenerate fields $(m,n)$. The resulting expressions, however, are quite complicated and will not be presented here.}
The
pentagon/hexagon identities involving one degenerate field produce
a recursion relation for the full, general fusion matrix which can, in principle, be used to determine
 its functional form.
 More prosaically, these identities are central to the results we present in this paper.

\bigskip

\section{Semi-classical conformal blocks for $N_f=4$}

Here we consider the conformal block corresponding to the four
punctured sphere, that describe the partition function of the $N_f=4$ theory.
We focus on the semiclassical limit, $\hbar \rightarrow 0$, with the conformal
dimensions given by $\Delta_i =\frac{Q^2}{4}+\frac{a_i^2}{\hbar}$ (where $a_i$
can be a mass parameter or a Coulomb branch parameter). In this limit\begin{equation}
\label{classic4}
\bigl\langle  V_{m_1}(0) V_{m_2}(q) V_{m_3}(1) V_{m_4}(\infty) \bigr\rangle_{a} \sim \exp \left(-\frac{{\cal F}(a)}{\hbar^2}+{\cal O}(\hbar^0) \right)
\end{equation}
As seen in Section 2, we can also consider the above conformal block with a degenerate field insertion. This insertion modifies the semi-classical limit (\ref{classic4}) at subleading order
\begin{equation}
\label{classicins}
\bigl\langle  V_{m_1}(0) V_{m_2}(q) \Phi_{2,1}(z) V_{m_3}(1) V_{m_4}(\infty) \bigr\rangle_{a} \sim \exp \left(-\frac{{\cal F}(a)}{\hbar^2}+\frac{{b \cal W}(a,z)}{ \hbar}+{\cal O}(\hbar^0) \right)
\end{equation}
Notice that we have inserted the degenerate field in $z$, with $q \ll z \ll 1$. Both expressions can be computed as an instanton expansion. (\ref{classic4}) simply as an expansion of the form $\sum_k {\cal Z}_k q^k$, while (\ref{classicins}) and an expansion of the form $\sum_{k,l} {\cal Z}_{k,l} q_1^k q_2^l$. In order to work consistently at a given order in the second sum, we choose $q_1=q^{1/2} z$ and $q_2=q^{1/2}/z$. This means we are locating the degenerate insertion at $q^{1/2} z$.

The superpotential ${\cal W}$ can then be obtained as the ratio of (\ref{classicins}) to (\ref{classic4}). To do this,
the correct procedure is then to expand its log powers of $q$ and then take the semiclassical limit. In first order we obtain
\begin{equation}
\label{superp}
{\cal W}(q^{1/2}z,a)=-a b^2 \log z+b^2 \frac{a^2+m_3^2-m_4^2-(a^2-m_1^2+m_2^2)z^2}{2 a z}q^{1/2}+...
\end{equation}
where the first term comes from a three level factor, which in the semiclassical limit is of the form $|z|^{-\frac{a b}{\hbar}}$.

As mentioned in Section 2, we can recover the quadratic differential $\phi_2(z)$ by considering the conformal block with an extra insertion of the energy momentum tensor. We consider the conformal block (\ref{classic4}) and insert an energy momentum tensor  $T(q^{1/2}z)$. Again, we can compute $\phi_2(q^{1/2}z)=-T(q^{1/2}z)$ as an instanton expansion. Considering its semiclassical limit
\begin{equation}
\label{quad}
\phi_2(q^{1/2}z)=\frac{a^2}{z^2}\frac{1}{q}-\frac{(m_1^2-m_2^2)z^2+m_4^2-m_3^2-a^2(1+z^2)}{z^3}\frac{1}{q^{1/2}}+...
\end{equation}
We obtain the following relation between (\ref{superp}) and (\ref{quad})

\begin{equation}
\left(\partial_z {\cal W}(q^{1/2}z) \right)^2=q \phi_2(q^{1/2}z)
\end{equation}

We checked this relation to rather high order in the instanton expansion.
This relation is exactly what we expect from the discussion of Section 2, see Eq. (\ref{difw}).

\section{Self-intersecting paths}

\begin{figure}[b] \centering \includegraphics[width=5.8in]{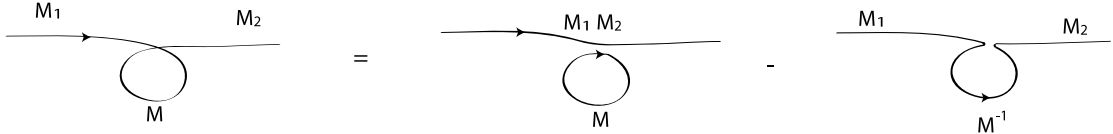} \caption{Decomposition of a self-intersecting path into a linear combination of non self-intersecting paths.} \label{selfintersecting}\end{figure}

We would like to argue that the Verlinde operators associated to self-intersecting paths can be rewritten as a linear combination of products of operators associated to non-self-intersecting paths. We provide a simple illustrative example. A full proof is outside the scope of this work.

Be $M_\gamma$ the operator valued $2 \times 2$ matrix which represents the
transport of a degenerate field along a path $\gamma$. If $\gamma$ is non-self-intersecting,
there always is a pair of pants decomposition of the surface such that $\gamma$ is one of the curves cutting the tubes. In that conformal block basis, $M_\gamma$ is a simple diagonal matrix $\Omega^2$, not operator valued, and is the core of a Wilson loop operator, written schematically as
$F M_\gamma F$. From the explicit computation, we know that the Wilson loop gives
$- \sec \pi b^2 \cosh 2 \pi b P_\gamma$. This is proportional to the trace ${\mathrm Tr} M_\gamma = - \exp \pi i b^2 \cosh 2 \pi b P_\gamma$. The determinant is ${\mathrm det} M_\gamma = - \exp 2 \pi i b^2$
Any $2 \times 2$ matrix satisfies the simple identity $M + M^{-1} \det M = Tr M$.
Hence we can write $M_\gamma = Tr M_\gamma + M_{\gamma}^{-1} \exp 2 \pi i b^2$.

Consider now a figure eight path $\gamma$, with a single self intersection point. If
we cut the path at the self-intersection, we decompose $\gamma$ into two non-self-intersecting, fragments, $\gamma_1$ and $\gamma_2$. There are now two possibilities. If $\gamma_1$ and $\gamma_2$ are
homotopic to each other, so that $\gamma = 2 \gamma_1$, then we can simply rewrite
$M_\gamma = M_{\gamma_1}^2 = M_{\gamma_1} Tr M_{\gamma_1} + \exp 2 \pi i b^2$,
and hence decompose the loop operator for $\gamma$ as a linear combination of the square of the loop operator for $\gamma_1$ and the identity. If $\gamma_1$ and $\gamma_2$ are not homotopic to each other,
we can pick a pair of pants decomposition where both $\gamma_1$ and $\gamma_2$ cut tubes.
Then both  $M_{\gamma_1}$ and $M_{\gamma_2}$ are actual matrices, not operator valued,
and we can rewrite $M_{\gamma_1} M_{\gamma_2} = M_{\gamma_1} Tr M_{\gamma_2} + M_{\gamma_1} M_{\gamma_2}^{-1} \exp 2 \pi i b^2$. Hence the loop operator for $\gamma$ is rewritten as a linear combination
of the product of loop operators for $\gamma_1$ and $\gamma_2$ and the loop operator for the path
$\gamma_1 \gamma_2^{-1}$, which is not self-intersecting.

The analysis for more general self-intersecting paths is probably more complicated.
If we could treat the operator valued transport matrices as normal matrices,
it is easy to replace each self intersections with linear combinations of the two possible ways to recombine
the path without self intersection. It possible that a judicious choice of pant decompositions may allow one to ignore the operator nature of the coefficients of the transport matrices. If this were not the case,
the operator ordering problems would give extra commutator terms. The commutators between
functions of $a$ and operators shifting $a$ by multiples of $\hbar b$ would be subleading
in the $\hbar \to 0$ limit. Hence the relations we seek would be valid at least in the undeformed
gauge theory.

\section{${\cal N} = 4$ SYM: an explicit check} \label{app:nisfourcheck}

Here we compare the results of Sect.~\ref{sec:line} with the explicit gauge theory expressions of the Wilson 
and 't Hooft loop operators, for the case of $SU(2)$ ${\cal N}=4$ SYM on $S^4$,  determined in \cite{Pestun:2007rz,Gomis:2009ir}. This provides an extra
check of our choice of overall normalization of the loop operators.

The ${\cal N}=4$ SYM partition function on $S^4$ is a product of an instanton sum, a classical contribtion, and a one-loop factor 
(which in this case happens to be trivial). The combined result reads \cite{Pestun:2007rz} 
\begin{equation}
\label{nisfourpart}
{\cal Z}^{S^4}_{{}_{{\cal N} =4}}=\, \Bigl| \frac{1}{\eta(q)}\Bigr|^2 \, \int_{\mathfrak{t}} \!  da\, \bigl| e^{2\pi i \tau a^2}\bigr|^2\, = \; \Bigl(\frac{1}{4 \tau_2}\Bigr)^{1/2} \Bigl| \frac{1}{\eta(q)}\Bigr|^2, 
\end{equation}
where $\eta(q)$ is the Dedekind $\eta$-function and with $q = e^{2\pi i \tau}$. The above gauge theory partition sum coincides with the
partition function of $c=25$ Liouville theory on $T^2$. It 
is invariant under $SL(2,\mathbb{Z})$ 
transformations, generated by $\tau \to \tau + 1$ and the S-duality map
\begin{equation}
\label{smap}
\tau \to \tilde{\tau} = -{1}/{\tau}\, .
\end{equation}

As shown in \cite{Pestun:2007rz}, the Wilson loop expectation value (normalized such that the Wilson loop in the trivial representation
equals to 1) takes the form
\begin{equation}
\label{nisfourwilson}
\Bigl\langle W_{j} \Bigr\rangle  =  \sqrt{4 \tau_2} \int_{\mathfrak{t}} \!  da\, \bigl| e^{2\pi i \tau a^2}\bigr|^2 \sum_{p=-j}^j e^{4\pi i p a}
 =   
%\; = \; 
\sum_{p=-j}^j e^{\frac{2\pi p^2}{\tau_2}}
\end{equation}
This result matches with our prescription in Sect.~5.1, based on the Verlinde loop operators of Liouville CFT on $T^2$.

We now compare the gauge theory result (\ref{nisfourwilson}) with the $m\to 0$ limit of 
our expression (\ref{thooftintegral}) for the 't Hooft loop  in the $\cN=2^*$ theory.
The action of the  't Hooft loop with general $j$ of the $\cN=4$ theory 
was determined in \eqref{generalthooft}, and indeed, $H_{1/2}$ of the $\cN=2^*$ theory \eqref{hooftl} 
reduces to \eqref{generalthooft} in the $m\to 0$ limit.
Then, adopting the same overall normalization as above, 
Eqs.~\eqref{generalthooft} and \eqref{thooftintegral} yield\footnote{Note that the DOZZ pre-factor (\ref{N2ZZform}) reduces to 1 in the $m\to 0$ limit.}
\begin{equation}
\label{nisfourthooft}
\Bigl\langle H_{j} \Bigr\rangle   =  \sqrt{4 \tau_2} \int_{\mathfrak{t}} \!  da\, e^{2\pi i\bar\tau a^2} \sum_{p=-j}^j 
e^{2\pi i \tau (a + p)^2 }
 =  \sum_{p=-j}^j e^{\frac{2 \pi p^2 |\tau|^2}{ \tau_2}}
\end{equation}
Here we used that the % ${\cal N}=4$ SYM 
chiral partition function is given by ${\cal Z}(a) = {e^{2\pi i \tau a^2} }/{\eta(q)}$,
and normalized the result by dividing by the $S^4$ partition sum (\ref{nisfourpart}), as prescribed. 
Eqn (\ref{nisfourthooft}) is manifestly S-dual to (\ref{nisfourwilson}), since 
%\begin{equation}
$1/\tilde{\tau}_2 = |\tau|^2/\tau_2$.

%%%%%%%%%%%%%%%%%%%%%%%%%%%%%%%%%%%%%%%%%%%%%%%%%%%%%%%%%%%%%%%%%%%%%%%

\bibliography{All}{}
\end{document}